\documentclass[conference,compsoc]{IEEEtran}
\AtBeginDocument{%
  }

\usepackage{amsmath,amsfonts}%
\usepackage{graphicx}
\usepackage{textcomp}
\usepackage{xcolor}
\usepackage{makecell}
\usepackage{multirow}

\usepackage{wasysym}
\usepackage{cite}

\usepackage[hyphens]{url}
\usepackage[bookmarks=true,breaklinks=true,colorlinks,linkcolor=black,citecolor=blue,urlcolor=black]{hyperref}

\usepackage{fancyhdr}
\usepackage[normalem]{ulem}
\usepackage{tikz}
\usepackage{pgfplots}
\usetikzlibrary{calc,positioning}
\usepackage{algorithm}
\usepackage[noend]{algorithmic}
\usepackage{soul}
\usepackage{xspace}
\usepackage{enumitem}
\usepackage{physics}
\usepackage{mdframed}
\usepackage{booktabs}

\usepackage{listings}
\setlist{nosep}
\setlist[itemize]{leftmargin=*}
\setlist[enumerate]{leftmargin=*}

\newif\iftight
\tighttrue

\newif\ifcomments
\commentstrue%

\newcommand{\revision}[1]{{#1}}

\newcommand{\sys}{\textsc{NanoTag}\xspace}
\newcommand{\sysns}{\textsc{NanoTag}}

\newcommand{\numbenchmarks}{10\xspace}
\newcommand{\arm}{ARM\xspace}
\newcommand{\mte}{MTE\xspace}
\newcommand{\mtens}{MTE}
\newcommand{\mtefull}{ARM's Memory Tagging Extension\xspace}
\newcommand{\device}{Google Pixel 8\xspace}
\newcommand{\devicemodel}{Google Pixel 8 Pro\xspace}
\newcommand{\scudo}{Scudo\xspace}

\newcommand{\scudofull}{Scudo Hardened Allocator\xspace}

\newcommand{\tagstore}{tag storage\xspace}
\newcommand{\addrtag}{address tag\xspace}
\newcommand{\memtag}{memory tag\xspace}
\newcommand{\tbi}{TBI\xspace}
\newcommand{\tbifull}{Top Byte Ignore\xspace}
\newcommand{\granule}{tag granule\xspace}
\newcommand{\tagcache}{tag cache\xspace}
\newcommand{\mtesync}{MTE SYNC mode\xspace}
\newcommand{\mteasync}{MTE ASYNC mode\xspace}
\newcommand{\mteasymm}{MTE ASYMM mode\xspace}

\newcommand{\exceptioninsn}{exception instruction\xspace}
\newcommand{\faultaddr}{fault address\xspace}
\newcommand{\android}{Android\xspace}
\newcommand{\androidns}{Android}
\newcommand{\ios}{iOS\xspace}
\newcommand{\oob}{buffer overflows\xspace}
\newcommand{\oobsingle}{buffer overflow\xspace}

\newcommand{\uaf}{use-after-free\xspace}

\newcommand{\geekbench}{Geekbench 6\xspace}

\newcommand{\juliet}{Juliet Test Suite\xspace}
\newcommand{\julietfull}{NIST Juliet Test Suite (v1.3)\xspace}
\newcommand{\termux}{Termux\xspace}
\newcommand{\asan}{Address Sanitizer\xspace}
\newcommand{\hwasan}{Hardware-assisted Address Sanitizer\xspace}

\newcommand{\taggranule}{tag granularity\xspace}
\newcommand{\taggranules}{tag granularities\xspace}
\newcommand{\intragranule}{intra-granule\xspace}

\newcommand{\shortgranule}{short granule\xspace}
\newcommand{\shortgranules}{short granules\xspace}

\newcommand{\Shortgranules}{Short granules\xspace}
\newcommand{\handler}{tag mismatch handler\xspace}

\newcommand{\tripwire}{tripwire\xspace}
\newcommand{\tripwirecap}{Tripwire\xspace}

\newcommand{\fault}{tag mismatch fault\xspace}
\newcommand{\boundscheck}{overflow detection\xspace}
\newcommand{\lastbyte}{the last byte\xspace}

\newcommand{\lastbits}{the last 4 bits\xspace}
\newcommand{\tripwiretag}{\tripwire tag\xspace}
\newcommand{\topbyte}{the topmost byte\xspace}
\newcommand{\violation}{memory safety bug\xspace}
\newcommand{\violationcap}{Memory safety bug\xspace}

\newcommand{\breakpoint}{breakpoint\xspace}
\newcommand{\valgrind}{Valgrind\xspace}
\newcommand{\retrowrite}{ASAN-Retrowrite\xspace}

\newcommand{\surface}{surface of bypassing \mte checks\xspace}
\newcommand{\spec}{SPEC CPU 2017 benchmarks\xspace}

\newcommand{\specint}{SPECrate Integer benchmarks\xspace}

\newcommand{\geomean}{geomean\xspace}
\newcommand{\vulmem}{vulnerable memory allocation\xspace}
\newcommand{\slowstart}{slow start\xspace}
\newcommand{\samplingphase}{sampling phase\xspace}
\newcommand{\alloccount}{\textit{AllocCount}\xspace}
\newcommand{\allocthreshold}{\textit{AllocThreshold}\xspace}
\newcommand{\samplingrate}{\textit{SamplingRate}\xspace}
\newcommand{\samplingratenospace}{\textit{SamplingRate}}
\newcommand{\acccount}{\textit{AccessCount}\xspace}
\newcommand{\accthreshold}{\textit{AccessThreshold}\xspace}
\newcommand{\trap}{trap\xspace}
\newcommand{\gwpasan}{GWP-ASan\xspace}
\newcommand{\magma}{Magma\xspace}
\newcommand{\aflpp}{AFL++\xspace}

\newcommand{\sotaadj}{state-of-the-art\xspace}

\newcommand{\julietoverflowscudo}{$75.68\%$\xspace}
\newcommand{\julietoverflowscudomissed}{$24.32\%$\xspace} %
\newcommand{\julietoverflowscudoasync}{$75.60\%$\xspace}
\newcommand{\julietoverflowasan}{$98.66\%$\xspace}
\newcommand{\julietoverflowscudolessthanasan}{$22.98\%$\xspace}

\newcommand{\julietoverflowsys}{$97.57\%$\xspace}
\newcommand{\julietoverflowsyslessthanasan}{$1.09\%$\xspace}

\newcommand{\julietdoublefreesys}{$98.37\%$\xspace}

\newcommand{\julietuafsys}{$96.69\%$\xspace}

\newcommand{\geekbenchasync}{$1.96\%$\xspace}
\newcommand{\geekbenchsync}{$3.76\%$\xspace}
\newcommand{\geekbenchsys}{$4.99\%$\xspace}
\newcommand{\geekbenchsysoversync}{$1.23\%$\xspace}
\newcommand{\geekbenchvalgrind}{$1348.60\%$\xspace}

\newcommand{\specshortcountratio}{$86.57\%$\xspace}

\newcommand{\specintasync}{$4.00\%$\xspace}
\newcommand{\specintsync}{$11.98\%$\xspace}
\newcommand{\specintsys}{$12.50\%$\xspace}
\newcommand{\specintasan}{$95.11\%$\xspace}

\newcommand{\specintsysmorethansync}{$0.52\%$\xspace}
\newcommand{\specintsysmorethansyncmax}{$6.13\%$\xspace} %
\newcommand{\specintsyncmorethanasync}{$7.98\%$\xspace}

\newcommand{\perlsys}{$47.60\%$\xspace}
\newcommand{\perlasan}{$182.59$\%\xspace}
\newcommand{\perlsyncmorethanasync}{$39.49$\%\xspace}

\newcommand{\handlerloc}{893\xspace} %
\newcommand{\scudomodifyloc}{227\xspace} %

\newcommand{\fuzzingsys}{15.86\%\xspace}
\newcommand{\fuzzingasan}{111.20\%\xspace}

\newcommand{\appsys}{12.35\%\xspace}
\newcommand{\appsyshs}{6.13\%\xspace}
\newcommand{\appasan}{153.9\%\xspace}

\newcommand{\ignore}[1]{}
\newcommand{\ie}{\textit{i.e.},\xspace}
\newcommand{\eg}{\textit{e.g.},\xspace}
\newcommand{\bench}[1]{\texttt{\small #1}}
\newcommand{\pgheading}[1]{\noindent\textbf{#1.}}

\newcommand{\observation}[2]{
\begin{mdframed}[nobreak=true,
  skipabove=0.5\topsep,
  skipbelow=0.5\topsep,
  linewidth=0.5mm
  ]
  \textbf{Observation:} #1\\
  \textbf{Insight:} \emph{#2}
\end{mdframed}
}

\newcommand{\takeaway}[1]{
\begin{mdframed}[nobreak=true,
  skipabove=0.5\topsep,
  skipbelow=0.5\topsep,
  linewidth=0.5mm
  ]
  \textbf{Takeaway:} #1
\end{mdframed}
}

\newcommand{\titletext}{\sys: Systems Support for Efficient Byte-Granular Overflow Detection on ARM MTE}

\title{\titletext}

\author{ 
    \IEEEauthorblockN{
    Mingkai Li\IEEEauthorrefmark{2},
    Hang Ye\IEEEauthorrefmark{2},
    Joseph Devietti\IEEEauthorrefmark{3},
    Suman Jana\IEEEauthorrefmark{2},
    Tanvir Ahmed Khan\IEEEauthorrefmark{2}
    }
    \IEEEauthorblockA{
    \IEEEauthorrefmark{2} Columbia University \hspace{0.2cm} \IEEEauthorrefmark{3} University of Pennsylvania
    \\
    }
    \IEEEauthorblockA{
    \{mingkai.li, hy2891\}@columbia.edu, devietti@cis.upenn.edu, suman@cs.columbia.edu, tk3070@columbia.edu \\
    }
}

\usepackage[available,functional]{ieeebadges}

\begin{document}
\pagenumbering{arabic} %

\maketitle

\begin{abstract}
Memory safety bugs, such as \oob and {\uaf}s, are the leading causes of software safety issues in production. Software-based approaches, \eg \asan (ASAN), can detect such bugs with high precision, but with prohibitively high overhead. \mtefull (\mte) offers a promising alternative to detect these bugs in hardware with a much lower overhead.  %
In this paper, we perform a thorough investigation of %
the first production implementation of ARM \mte (\device) and observe
that 
\mte can only achieve coarse precision in bug detection compared with software-based approaches such as ASAN, mainly due to its 16-byte \taggranule.

\revision{To address this issue, we present \sys\footnote{We open-source our work at \url{https://github.com/ice-rlab/NanoTag}.}, a system to probabilistically detect \oob at byte granularity in unmodified \mte-enabled binaries with minimal changes to memory allocators, introducing an explicit detection-performance tradeoff for in-house testing.}
\sys detects \oob at byte granularity by setting up a \tripwire for {\granule}s that may require \intragranule overflow detection. The memory access to the \tripwire causes additional \boundscheck in the software while 
using \mte's hardware to detect bugs
for the rest of the accesses. We implement \sys based on the \scudofull, the default memory allocator on \android since \android 11. Our evaluation results across popular benchmarks and real-world case studies show that \sys detects nearly as many {\violation}s as ASAN while incurring similar run-time overhead to \scudofull in \mtesync.

\end{abstract}

\section{Introduction}
\label{sec:intro}

{\violationcap}s, such as buffer overflows and {\uaf}s, remain the dominant causes of software vulnerabilities---70\% of Microsoft~\cite{mattmiller} and 51\% of Android~\cite{aospmemsafety} vulnerabilities.
Consequently, researchers have proposed a wide range of static and dynamic techniques over the years~\cite{wang2012improving, li2023hybrid, zhang2025statically, bae2021rudra, facebookinfer, asan, asan--, duck2018effectivesan, giantsan, zhang2021sanrazor, yu2024shadowbound, nagarakatte2009softbound, hwasan}.
Among these techniques, memory safety sanitizers~\cite{asan, asan--, duck2018effectivesan, giantsan, zhang2021sanrazor, yu2024shadowbound, nagarakatte2009softbound, hwasan} 
provide high detection accuracy, as they observe the executions of applications directly. %

Unfortunately, sanitizers incur significant run-time overhead.
\asan~\cite{asan} and \hwasan~\cite{hwasan} impose $\sim$$2\times$ performance overhead~\cite{overheadasan, hwasan}. The overhead of binary instrumentation is even higher---$3\times$ for \retrowrite~\cite{dinesh2020retrowrite}, $17\times$ for Valgrind~\cite{nethercote2007valgrind}, and $35\times$ for QASAN~\cite{fioraldi2020qasan}.
During in-house testing, such as fuzzing, reduced throughput means sanitizers cannot run continuously. %
High-throughput fuzzers skip sanitization for most tests and %
re-run a reduced corpus with sanitization~\cite{sanddecouple}. Skipping sanitization for most tests stops fuzzers from detecting silent memory safety bugs~\cite{fuzzan}.
Detecting such silent bugs during in-house testing requires efficient sanitization.

\pgheading{MTE} Hardware vendors aim to make sanitization efficient %
by introducing mechanisms such as \mtefull (\mte).
\mte maintains tags for pointers and memory bytes. Matching a pointer's tag against the tag of any 16-byte memory the pointer accesses directly in hardware, \mte sanitizes memory accesses without heavy instrumentation.
ARM reports only 1–2\% overhead for the fastest \mteasync~\cite{blogarm}\footnote{\mte supports SYNC, ASYNC, and ASYMM modes; see \S\ref{sec:analysis-sub-hardware}.}.
\device is the first implementation of \mte. %
Google also added support for \mte to \scudofull~\cite{androidscudo}, \androidns's default memory allocator, reporting no noticeable slowdown~\cite{projectzero}.
\mte has already helped catch real-world issues such as Google CVE-2024-23694~\cite{grapheneosmte}.
Consequently, \mte is one of the few sanitization techniques potentially usable continuously for in-house testing and fuzzing~\cite{sync}.

\pgheading{\mte's limitations}
In this paper, we conduct a thorough investigation of \device to show
that \mte's low performance overhead comes with a significant loss in its bug detection capability.
In particular, we find that \mte misses \julietoverflowscudomissed of heap-based buffer overflows in \juliet~\cite{juliet}, whereas ASAN detects %
\julietoverflowasan
of such overflows.
The root cause is architectural: \mte assigns a 4-bit tag to each 16-byte memory, \textit{\granule}, and checks tags only at that granularity.
\mte fails to detect any overflow that %
occurs within a single \granule, even when it overwrites useful data~\cite{gnuallocator} or padding~\cite{androidscudo}, based on memory allocators.
We refer to these silent cases as \emph{\intragranule \oob}.

These \intragranule \oob have significant implications
in testing and fuzzing~\cite{sync}, where the goal is to detect silent latent bugs. %
Such in-house testing may tolerate~\cite{sanddecouple} 
up to 20\% overhead of some \mte modes~\cite{stickytags}.
However, the loss in \mte's detection capability significantly undermines the effectiveness of in-house testing.
Overflows that appear benign (\eg accessing padding bytes) may become exploitable under different inputs or memory layouts~\cite{khan2019huron} that place critical data within the same 16-byte region.
Thus, undetected \intragranule \oob can directly lead to latent security vulnerabilities.

Quantitatively, in \S\ref{sec:analysis}, we show that a significant fraction (\specshortcountratio) of memory allocations in the \spec suite~\cite{specbenchmark} allow potential \intragranule \oob. %
Such overflows also %
cause real-world CVEs, including CVE-2024-12084~\cite{cveexample}, rated critical by Google and Red Hat~\cite{securitygoogle, redhatrating}.
Mitigating these issues 
requires byte-granular overflow detection, which %
\mte %
cannot provide.

\pgheading{Challenges} %
Byte-granular overflow detection is challenging due to
significant 
run-time and memory overhead.
Detecting such overflows in software requires additional instrumentation similar to existing sanitizers~\cite{asan, hwasan, dinesh2020retrowrite, nethercote2007valgrind, fioraldi2020qasan}, adding at least $2\times$ run-time overhead~\cite{overheadasan}. Moreover, additional metadata in the form of redzones~\cite{hastings1992purify, nethercote2007valgrind, efence} or shadow memory~\cite{nethercote2007valgrind} increases memory usage by at least 10–35\%~\cite{hwasan}.
In terms of memory overhead, hardware support is even more expensive as reducing \mtens’s granularity from 16~bytes to 1~byte would require storing 4~bits of metadata per byte, consuming one-third of physical memory.

\pgheading{Our Solution} 
\revision{In this paper, we present \sys, the first system to detect {\violation}s probabilistically at byte granularity on real ARM \mte hardware, addressing \intragranule \oob in unmodified \mte-enabled binaries.}
During allocation, \sys identifies {\granule}s that permit \intragranule \oob, which we call \emph{\shortgranule}s.
For these \shortgranules, \sys probabilistically enables byte-level checking in software while retaining \mtens’s native coarse-granular detection of \oob and {\uaf}s in hardware.
\revision{In particular, \sysns’s probabilistic approach introduces an explicit tradeoff between bug detection capability and performance overhead through a controllable sampling knob. Such a knob enables \sysns’s primary use cases, \ie in-house testing (\eg fuzzing), to select the appropriate balance between detection strength and overhead.}
As a result, \sys adds only minimal overhead on top of \scudofull in \mtesync.

\sys avoids high run-time overhead by relying primarily on hardware checks.
It inserts additional software checks only while accessing a \shortgranule. %
\sys sanitizes such accesses via a \emph{\tripwire}: a \granule tagged with a value intentionally %
different
from the pointer's tag.
\sys installs {\tripwire}s probabilistically on \shortgranules.
While accessing a \shortgranule with a \tripwire,
hardware raises a \fault, triggering \sysns’s software handler.
The handler checks for \oobsingle. %
If the access is benign, \sys resumes normal execution.
For violations, \sys reports detailed diagnostic information, including tag and register values.

\sys avoids memory overhead by storing metadata inside unused padding bytes within each \shortgranule and 
inferring access offsets from \mtesync information.
Specifically, \mtesync provides the \exceptioninsn and \faultaddr, enabling \sys to compute the exact byte offset of the access.
The padding bytes inside a \shortgranule contain its \emph{real tag}. %
The tag \sys assigns to the granule to install a \tripwire, instead, encodes the number of valid addressable bytes.
Upon a \fault, \sys extracts both tags from the granule, %
verifies that the real tag matches the pointer's tag, and confirms that the access is within the addressable region.

If the access is valid, \sys temporarily removes the \tripwire by restoring the granule’s tag, sets a \emph{\trap} (e.g., hardware or software breakpoint~\cite{breakpoint}) immediately after the \exceptioninsn, and replays the instruction without raising another fault.
When the program hits the trap,
\sys reinstalls the \tripwire and records how frequently it is triggered.
Once a \tripwire exceeds a threshold, \sys removes it permanently to avoid unnecessary overhead.

We implement \sys on top of \scudo and evaluate it systematically on \device.
\sys detects {\violation}s at byte granularity while maintaining runtime overhead comparable to \scudo in \mtesync.
On \juliet, \sys detects \julietoverflowsys heap-based overflows, closely matching ASAN (\julietoverflowasan) and significantly outperforming both \mte SYNC (\julietoverflowscudo) and ASYNC (\julietoverflowscudoasync).
On \specint, \sys incurs a \geomean overhead of \specintsys, similar to \scudo in \mtesync (\specintsync).
\revision{In real-world case studies, \sys incurs \geekbenchsys overhead on closed-source \geekbench~\cite{geekbench}, %
up to \appsys overhead on three large open-source real-world applications (\bench{Memcached}~\cite{memcached}, \bench{LevelDB}~\cite{leveldb}, \bench{RocksDB}~\cite{rocksdb}), and slows fuzzing throughput by only \fuzzingsys across three \magma targets—one-seventh of ASAN’s slowdown (\fuzzingasan).}

\pgheading{Contributions} This paper contributes %

\begin{itemize}
    \item A thorough study of ARM \mte's hardware and software stack on \device to quantitatively show that \intragranule \oob expose a large \surface (\specshortcountratio) in \spec.
    \item \revision{\sys, a system to efficiently detect {\violation}s probabilistically in unmodified \mte-enabled binaries at byte granularity, addressing \intragranule \oob in real hardware for the first time to help in-house testing detect such bugs with an explicit detection-performance tradeoff.}
    \item An evaluation of \sys's prototype on \juliet and \spec, to show that \sys achieves similar bug detection capability (\julietoverflowsys) as ASAN (\julietoverflowasan) and incurs a run-time overhead of \specintsys, which is
    $7\times$
    lower than the %
    \specintasan%
    overhead of ASAN and comparable to the \specintsync overhead of the \mtesync.  %
\end{itemize}

\section{Characterizing \mtefull on \device}
\label{sec:analysis}

\begin{figure}[t]
    \centering
    \includegraphics[width=\linewidth]{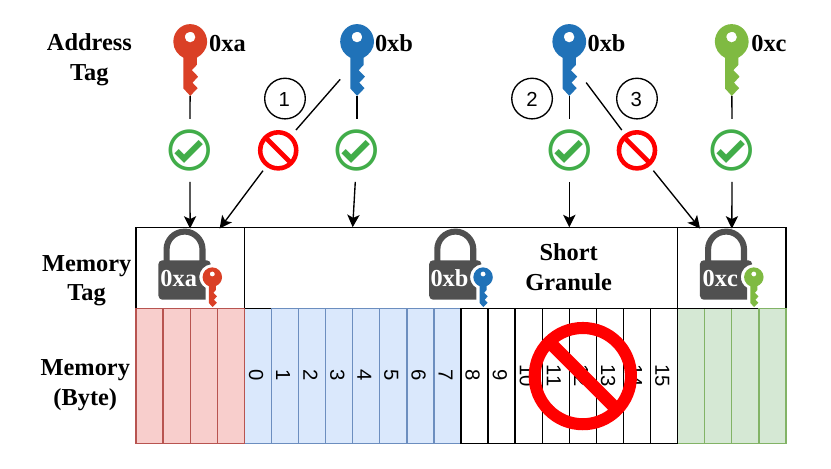}
    \caption{\small An overview of ARM \mte. Locks and keys represent %
    memory and address tags, respectively. %
    Addressable bytes are colored. The middle \granule %
    is a \shortgranule. \mte detects both underflow (\bench{1}) and overflow (\bench{3}) across the \granule boundary. However, it %
    ignores the \intragranule \oobsingle (\bench{2}).
    }
    \label{fig:mte-overview}
\end{figure}

\subsection{Background}
\label{sec:analysis-mte-stack}

We first describe ARM \mte's hardware and then provide a brief summary of the relevant software on \device.

\subsubsection{Hardware} \label{sec:analysis-sub-hardware} In a nutshell, \mte detects memory safety bugs using a lock-and-key approach~\cite{whitepaper}, as we show in Fig.~\ref{fig:mte-overview}. While accessing memory bytes with a pointer, the pointer carries a \textit{key} associated with a \textit{lock} that protects the memory bytes. \mte permits the memory access only if the pointer's key matches the memory bytes' lock. Otherwise, \mte generates an exception to report a \violation.

\pgheading{\mte Tag Management} 
ARM \mte implements these locks and keys with \emph{tags}.
A tag is a 4-bit token stored either inline at the pointer or at a carved-out \textit{\tagstore} in the memory~\cite{mteefficient}. The tag that is inline at the pointer is called the \textit{\addrtag} (the key), storing at the leftmost byte of the pointer.
MTE does not treat %
it as part of the virtual address with the \tbifull (\tbi) feature~\cite{tbi}.
The tag that is at the carved-out \tagstore is called the \textit{\memtag} (the lock).
\mte divides the physical address space into 16-byte chunks called \textit{\granule},
associating a \memtag in the \tagstore with it.
There is also a dedicated \textit{\tagcache} for {\memtag}s~\cite{tagstore}. 
While bringing memory bytes to a processor's cache, ARM \mte also prefetches the corresponding \memtag from \tagstore to \tagcache~\cite{tagstore}.

\pgheading{\mte Tag Check} For each memory access, \mte compares the {\memtag} in the \tagstore (the lock) with the \addrtag of the pointer (the key). 
\mte permits the memory access only if \addrtag and all {\memtag}s involved in the memory access are the same. Otherwise, the processor raises a \textit{\fault} either synchronously or asynchronously.

\pgheading{\mte Modes} \mte modes determine when \mte reports the \fault. \mte currently supports 3 different modes~\cite{mtemodes}: (1) SYNC, (2) ASYNC, and (3) ASYMM. In \mtesync, the tag mismatch causes a synchronous exception immediately, while also recording the precise location of the \textit{\exceptioninsn} and the \textit{\faultaddr}. In \mteasync, the \fault causes an exception at the nearest kernel entry, \eg a system call or a timer interrupt. The processor continues execution after a \fault until the nearest kernel entry, recording only an imprecise location near the \exceptioninsn. \arm also introduces \mteasymm %
to handle read and write memory accesses in SYNC and ASYNC modes, respectively. However, \android does not allow users to manually enable \mteasymm. Instead, \mteasync is redefined as \mteasymm in compatible devices~\cite{mtemodes}.

\begin{table}[t]
    \centering
    \caption{Instructions in ARM \mte.}
    \begin{tabular}{p{0.4\linewidth}|p{0.5\linewidth}}
    \hline
        \textbf{Usage} & \textbf{Instructions}  \\\hline
        Random tag generation & \bench{IRG}, \bench{GMI}\\\hline
        Pointer arithmetic & \bench{ADDG}, \bench{SUBG}, \bench{SUBP}\\\hline
        Tag load and store & \bench{LDG}, \bench{STG}, \bench{STZG}, \bench{ST2G}, \bench{STZ2G}, \bench{STGP}\\\hline
        Tag load and store (privileged) & \bench{LDGM}, \bench{STGM}, \bench{STZGM}\\\hline
        Data cache operations & \bench{DC GVA}, \bench{DC GZVA}, \bench{DC IGDSW}, \bench{DC IGDVAC}, \bench{DC IGSW}, \bench{DC IGVAC}\\\hline
    \end{tabular}
    \label{tab:mte-instruction}
\end{table}

\subsubsection{Instruction Set Architecture}
Once both {\addrtag}s and {\memtag}s are properly set up, ARM \mte detects {\violation}s for every load and store instruction~\cite{loadstoreinsns} by comparing the tags in the hardware~\cite{whitepaper}. To manage tags, \mte adds several instructions that we list in Table~\ref{tab:mte-instruction}.
In particular, \mte provides \bench{IRG} and \bench{GMI} instructions to generate a random tag with a certain mask. 
While generating random tags, \mte excludes values in the mask.
To set up the \addrtag, \mte provides pointer arithmetic instructions, \eg \bench{ADDG} and \bench{SUBG}. An instruction tags an untagged pointer either directly from a generated random tag or based on an existing tagged pointer.
To set up the \memtag, \mte provides tag load and store instructions, \eg \bench{LDG} and \bench{STG}, to access the \tagstore with the virtual address of the \granule. When storing to the \tagstore, instructions can choose to zero (\eg \bench{STZG}) or initialize (\eg \bench{STGP}) the corresponding \granule. A single user-space instruction could set up the \memtag for at most 2 {\granule}s~\cite{whitepaper}. Setting up the \memtag for multiple {\granule}s (maximum 16) requires instructions that the processor can execute only in the privileged mode ($\textit{Exception Level (EL)} \ge 1$). Along with these \mte instructions that prior works have studied~\cite{whitepaper, mteefficient, mtsan, stickytags}, we find that ARM also introduces data cache operations that allow setting up the \memtag for multiple {\granule}s in the user space. For example, Google uses the \bench{DC GZVA} instruction in \scudo to set up the {\memtag}s of the whole cache line and zero its content~\cite{scudodccommit}.

\subsubsection{Software} 
\device has end-to-end support for \mte in software, including Instruction Set Architecture~\cite{whitepaper}, operating system~\cite{kernelmte}, and \scudofull~\cite{androidscudo}.
Briefly, 
the boot loader carves out around 3\% of the physical memory for \tagstore at the system start up~\cite{bootloader}. The \android operating system supports configuring \mte
via system calls~\cite{kernelmte} and Android Debugging Bridge~\cite{projectzero}. %
\android's default memory allocator, \scudo~\cite{androidscudo}, uses \mte to detect {\violation}s for memory allocations by its primary allocator~\cite{mteefficient}. %

\subsection{How does \mte compare against purely software-based techniques?}
\label{sec:analysis-security-measurement}

In this section, we characterize ARM \mte's bug detection capability on \device %
using \julietfull~\cite{juliet}, similar to prior work~\cite{giantsan, li2022pacmem, stickytags, asan--, mtsan}.
Among different Common Weakness Enumerations (CWEs) of \juliet,
out-of-bound write, out-of-bound read, and use after free are ranked 2nd, 6th, and 8th of the top 25 most dangerous software weaknesses in 2024~\cite{cwe2024}.
We select the corresponding heap-related CWEs: %
\bench{CWE122} (heap-based buffer overflow), \bench{CWE415} (double free), and \bench{CWE416} (use after free). %
Each CWEs include both benign %
and buggy test cases. As benign test cases do not include any \violation, %
a sound bug detector should not report false positives. The buggy test cases include one or more {\violation}s. %
Similar to prior work~\cite{giantsan}, we exclude some \bench{CWE122} test cases to avoid infinite waiting and nondeterministic results: %
(1) \bench{CWE129\_listen\_socket} and \bench{CWE129\_connect\_socket} requiring external inputs; (2) \bench{CWE129\_rand} triggering bugs randomly. %
Due to \mte's probabilistic nature, we run each CWE with Scudo %
multiple times to report the average number.
For comparison, we run \juliet with %
ASAN~\cite{asan}, \gwpasan~\cite{serebryany2024gwp}, and \scudo in \mte SYNC and ASYNC modes. For ASAN, we disable Leak Sanitizer~\cite{lsan} to ensure that ASAN only detects \oob and {\uaf}s.
For \gwpasan, we set its sampling rate to the default value (5000). %
Due to sampling's probabilistic nature,
we also run each CWE multiple times with \gwpasan.%

\begin{table}[t]
    \centering
    \caption{\small Heap-based {\violation}s in \juliet. \scudo only detects 75\% of %
    buffer overflow bugs in \mte SYNC and ASYNC modes, while ASAN detects almost all (\julietoverflowasan) of them.
    }
    \label{tab:juliet-bad-measurement}
    {
    \footnotesize%
    \begin{tabular}{c|c|c|c|c|c}
    \hline
        \textbf{CWE ID} & \textbf{Total} & \textbf{ASAN} & \textbf{\makecell[c]{GWP-\\ASan}} & \textbf{\makecell[c]{Scudo*\\ (ASYNC)}} & \textbf{\makecell[c]{Scudo*\\ (SYNC)}}\\
    \hline
        \makecell[c]{\bench{\footnotesize CWE122}\\(heap-\\based\\ buffer\\ overflow)} & 3438 & 98.66\% & \makecell[c]{23.15\% \\± 0.24\%} & \makecell[c]{75.60\% \\± 0.13\%} & \makecell[c]{75.68\% \\± 0.10\%}\\\hline
        \makecell[c]{\bench{\footnotesize CWE415}\\(double\\ free)}                & 818  & 100\% & \makecell[c]{98.60\% \\± 0.23\%} & \makecell[c]{98.43\% \\± 0.14\%} & \makecell[c]{98.45\% \\± 0.19\%}\\\hline
        \makecell[c]{\bench{\footnotesize CWE416}\\(use\\ after\\ free)}             & 393  & 100\% & \makecell[c]{0\%} & \makecell[c]{96.54 \\± 0.60\%} & \makecell[c]{96.94\% \\± 0.58\%}\\
    \hline
    \end{tabular}
    }\\
    *The variance comes from \mte's tag collision across multiple runs.
\end{table}

\pgheading{Results} In our experiment, %
ASAN, \gwpasan, and \scudo do not report any of the benign test cases as bugs.
Therefore, we only show the results of buggy test cases in Table~\ref{tab:juliet-bad-measurement}.
Table~\ref{tab:juliet-bad-measurement} shows the percentages of {\violation}s ASAN, \gwpasan, and \scudo detect. 
As we show, ASAN detects almost all {\violation}s.
In comparison, \gwpasan falls significantly short as %
it detects {\violation}s in sampled memory allocations for %
in-production deployment.%

For \scudo, it has almost the same results in \mte SYNC and ASYNC modes. We show that, although \scudo with \mte support detects most of the temporal memory safety bugs, \ie \bench{CWE415} (double free) and \bench{CWE416} (use after free), it detects significantly fewer spatial memory safety issues, \ie \bench{CWE122} (heap-based buffer overflow), than ASAN.

For \oob (\bench{CWE122}), \scudo detects only \julietoverflowscudo and \julietoverflowscudoasync of heap-based buffer overflow bugs in \mte SYNC and ASYNC modes, while ASAN detects almost all (\julietoverflowasan) of them.
By analyzing the bugs \scudo fails to detect, we find that \scudo is less precise than ASAN, mainly due to its 16-byte \taggranule. The missed bugs in \bench{CWE122} are mostly \intragranule \oob, \ie a \oobsingle that occurs within the 16-byte \granule. Fig.~\ref{fig:mte-overview} shows an example of such an \intragranule \oobsingle. For access 2 in Fig.~\ref{fig:mte-overview}, it accesses the 12th byte in the \granule, while only the first 8 bytes 
are addressable. 
Therefore, access 2 results in an \intragranule \oobsingle. 
However, due to \mte's coarse \taggranule, \mte fails to detect such an \intragranule \oobsingle, as \mte permits accesses to the full \granule even though only part of it is addressable.

\observation{\mte detects only up to \julietoverflowscudo heap-based buffer overflow bugs on \juliet, while ASAN detects almost all (\julietoverflowasan) of them.
}{
\mte ignores \intragranule \oob due to 16-byte \taggranule.
}

\begin{figure}[t]
    \centering
    \includegraphics[width=\linewidth]{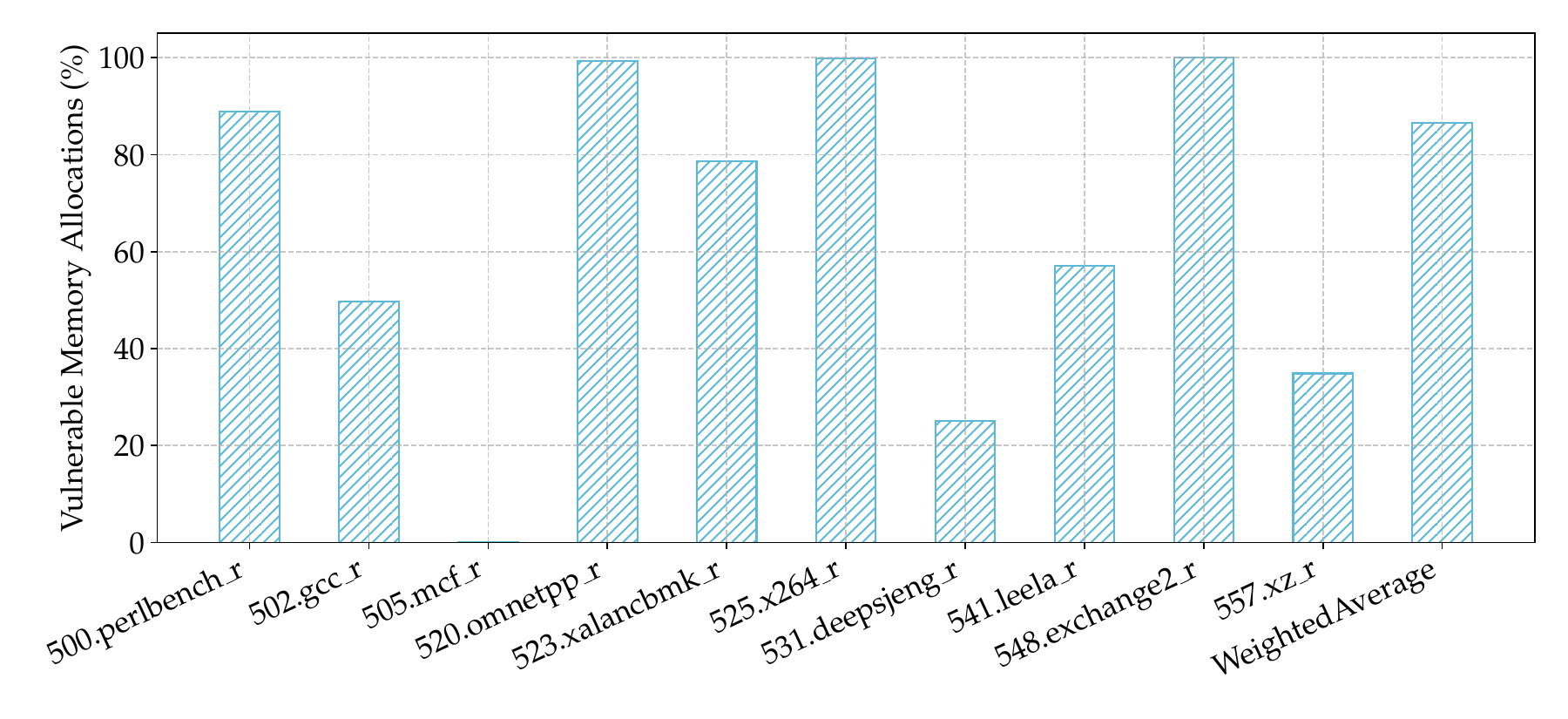}
    \caption{\small Percentage of {\vulmem}s in \spec. %
    \specshortcountratio of all allocations %
    in \spec include \shortgranules, leading to potential \intragranule \oob. The only outlier is \bench{505.mcf\_r} (0.05\%).
    }
    \label{fig:spec-dist}
\end{figure}

\subsection{How frequently do programs allocate \shortgranules?} %
\label{sec:analysis-distribution}

To understand the \surface due to missed \intragranule \oob, we analyze the memory allocations in \spec. Our experiment setup of \spec is consistent with our evaluation, which is described in \S\ref{sec:eval-setup} and \S\ref{sec:eval-performance}. A memory allocation is a \textit{\vulmem}
if it contains a \shortgranule.
We calculate the percentage of {\vulmem}s as the number of memory allocations, whose size is not a multiple of 16 bytes, divided by the total number of memory allocations in the benchmark. %

\pgheading{Results} As shown in Fig.~\ref{fig:spec-dist}, \specshortcountratio of all memory allocations in \spec include \shortgranules, leading to potential \intragranule \oob.
Specifically, 
70\% and 50\% of benchmarks have more than 40\% and 60\% vulnerable memory allocations, respectively. %
We also observe that there are three benchmarks (\bench{520.omnetpp\_r}, \bench{525.x264\_r}, and \bench{548.exchange2\_r}) with more than 99\% vulnerable memory allocations. 
This illustrates the ubiquitousness of {\vulmem}s in various applications.
One potential approach to reduce the ubiquity of these {\vulmem}s
would be to use a smaller \taggranule than 16 bytes. Unfortunately, such an approach %
would increase the size of the \tagstore, carving out more physical memory, %
$\frac{1}{3}$ of it in the extreme case (1-byte \taggranule). For example, while studying the impact of different \taggranules, prior work~\cite{kostyamemtag} argued that the size of an optimal \taggranule is 16 bytes as a larger \taggranule significantly increases the memory overhead due to alignment.

\observation{\specshortcountratio of all memory allocations in \spec include {\shortgranule}s.%
}{
\Shortgranules, leading to \intragranule \oob \mte ignores, are ubiquitous in allocations.
}

\begin{figure}[t]
    \centering
    \small
\begin{lstlisting}
#if defined SHA512_DIGEST_LENGTH
#define MAX_DIGEST_LEN SHA512_DIGEST_LENGTH
#elif defined SHA256_DIGEST_LENGTH
#define MAX_DIGEST_LEN SHA256_DIGEST_LENGTH
#elif defined SHA_DIGEST_LENGTH
#define MAX_DIGEST_LEN SHA_DIGEST_LENGTH
#else
#define MAX_DIGEST_LEN MD5_DIGEST_LEN
#endif
\end{lstlisting}
\bench{md-defines.h}, adapted from~\cite{mddefinesh}
\begin{lstlisting}
struct sum_buf {
  OFF_T offset;         /*offset in file of this chunk*/
  int32 len;            /*length of chunk of file*/
  uint32 sum1;          /*simple checksum*/
  int32 chain;          /*next hash-table collision*/
  short flags;          /*flag bits*/
  char sum2[SUM_LENGTH];/*checksum*/
};
\end{lstlisting}
\bench{rsync.h}, adapted from~\cite{codersynch}
\begin{lstlisting}
struct sum_struct *s = new(struct sum_struct);
// . . .
s->sums = new_array(struct sum_buf, s->count);
for (i = 0; i < s->count; i++) {
    s->sums[i].sum1 = read_int(f);
    read_buf(f, s->sums[i].sum2, s->s2length);
// . . .
\end{lstlisting}
\bench{sender.c}, adapted from~\cite{senderc}

\caption{\small CVE-2024-12084~\cite{cveexample}, an \intragranule heap-based buffer overflow bug in \textit{rsync} that becomes a silent data corruption in \mte, while both ASAN and \sys detect it.
}
\label{fig:cve-example}
\end{figure}

\subsection{How do %
\intragranule \oob lead to undetected real-world vulnerabilities?}
\label{sec:analysis-example}

In Fig.~\ref{fig:cve-example}, we show the code snippets of a real-world vulnerability, CVE-2024-12084~\cite{cveexample}. %
This vulnerability is due to a heap-based buffer overflow bug in \textit{rsync}~\cite{rsync}.
In particular, the buffer overflow occurs at line 6 of \bench{sender.c}, while filling up \bench{s->sums[i].sum2} with client-provided \bench{s->s2length} bytes~\cite{securitygoogle}. As shown in \bench{rsync.h}, \bench{sum2} is an array of \bench{SUM\_LENGTH} bytes, which is a field of a 40-byte structure, \bench{sum\_buf}. The attacker exploits this vulnerability by controlling the value of \bench{s->s2length}, whose maximum value \bench{MAX\_DIGEST\_LEN} is typically larger than \bench{SUM\_LENGTH}, as defined in \bench{md-defined.h}. As the size of \bench{sum\_buf} (40 bytes) is not a multiple of 16 bytes, the last \granule in the memory allocation has 8 non-addressable bytes. Therefore, \mte ignores any overflow into these 8 non-addressable bytes in the last \granule. %

We implement a simple Proof-of-Concept (PoC) of CVE-2024-12084 by setting \bench{s->s2length} to 24, initiating an \intragranule \oobsingle in the last \granule of \bench{sum\_buf}. 
As we evaluate this PoC with \mte-enabled \scudo, we observe that
\mte fails to detect the overflow in both SYNC and ASYNC modes due to its 16-byte \taggranule.
On the other hand, both ASAN and \sys prevent this PoC by detecting \intragranule \oob.

\observation{
Ignoring \intragranule \oob, \mte fails to detect a PoC of CVE-2024-12084.
}{Detecting \intragranule \oob helps uncover real-world vulnerabilities.
}

\section{Design}
\label{sec:design}

\label{sec:design-overview}

\begin{figure}[t]
    \centering
    \includegraphics[width=0.9\linewidth]{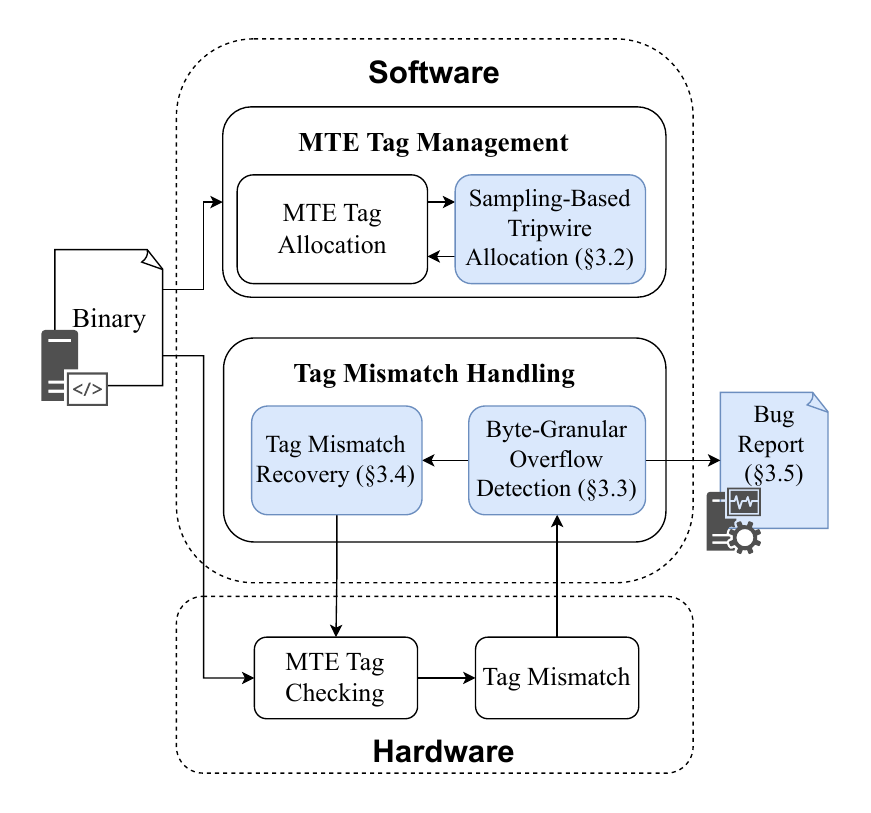}
    \caption{\small \sys's workflow. Blue blocks specify additional components in \sys, while white blocks denote unmodified components in \mte's hardware and software.
    }
    \label{fig:sys-workflow}
\end{figure}

\revision{In Fig.~\ref{fig:sys-workflow}, we show \sys's workflow to detect {\violation}s in unmodified \mte-enabled binaries at byte granularity.}
With an overview of \sys's design goals %
in \S\ref{sec:design-usage-model}, we describe its components in \S\ref{sec:design-short-granule}-\S\ref{sec:design-bug-report}.%

\subsection{Design Goals}
\label{sec:design-usage-model}

\sys
aims %
to detect {\violation}s in unmodified binaries during in-house testing. %
If the application binary contains {\violation}s, \ie \oob or {\uaf}s, 
\sys's %
goal is to detect these bugs:
\textbf{(D1)} %
at byte granularity, %
\textbf{(D2)} %
without instrumenting the source code or the binary, and %
\textbf{(D3)} %
with run-time overheads %
similar to
\mte-enabled \scudo. %
With these design goals, \sys ensures that in-house testing, \eg fuzzing campaigns, 
(1) does not ignore potential in-production vulnerability because of overflows to padding bytes, (2) detects as many bugs as possible, and (3) runs as fast as possible.
\sys achieves these design goals with two layers of sanitizations. For the first layer of sanitization, \sys detects \oob at 16-byte granularity and {\uaf}s for all memory allocations with ARM \mte's hardware checks. For the second layer of sanitization, \sys detects \oob at byte granularity, including \intragranule \oob, for sampled memory allocations (\S\ref{sec:design-short-granule}) with \sys's additional software checks (\S\ref{sec:design-bounds-check}, \S\ref{sec:design-recovery}). If \sys detects {\violation}s based on software and hardware checks, it reports them to application developers. The developer then analyzes the bug report (\S\ref{sec:design-bug-report}) to fix these bugs before deploying the application in production. %
\revision{We assume micro-architectural attacks against \mte~\cite{kim2025tiktag, stickytags, kocher2020spectre, lipp2020meltdown} and adversarial attacks to bypass byte-level checks are outside the scope of \sys's primary use case, in-house testing.}

\subsection{Sampling-Based Tripwire Allocation}
\label{sec:design-short-granule}

\sys avoids instrumenting the source code or the binary (D2) by setting up {\tripwire}s that invoke additional software checks. \sys sets up these {\tripwire}s
via sampling, amortizing the overhead of additional software checks (D3).

\begin{figure}[t]
    \centering
    \small
    \includegraphics[width=\linewidth]{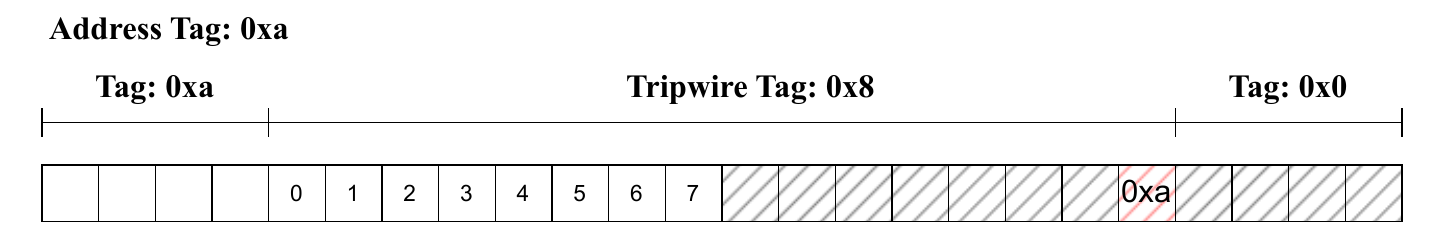}
    \caption{\small An example of how \sys assigns tags to {\shortgranule}s where only a fraction of 16 bytes is addressable (the middle one). Blocks in hatches indicate non-addressable bytes. We highlight the last byte of the \shortgranule in red.}
    \label{fig:tripwire}
\end{figure}

\pgheading{\tripwirecap} 
As we show in Fig.~\ref{fig:mte-overview}, 
a \textit{\shortgranule} is a \granule where only part of its 16 bytes is addressable,
\eg the \granule in the middle of Fig.~\ref{fig:mte-overview}. %
To detect \intragranule \oob in the \shortgranule, \sys needs to differentiate memory accesses to the \shortgranule from other granules. Therefore, \sys probabilistically sets up %
a \textit{\tripwire} for such {\shortgranule}s, \ie \sys allocates the \shortgranule a special \textit{\tripwiretag}, which is different from other tags in the same memory allocation. %
Similar to HWASAN~\cite{hwasan}, \sys sets the \tripwiretag as the number of addressable bytes in the \shortgranule, which is \bench{0x8} %
for the \shortgranule (the middle one) in %
Fig.~\ref{fig:tripwire}.
Consequently, when a pointer accesses the \tripwire, it always causes a \fault, 
either from a true \violation or only from a safe program access to the \shortgranule's addressable bytes,
which \sys intercepts in its \handler (\S\ref{sec:design-bounds-check}). %
To identify the root cause of the \fault, \sys stores the legit %
\addrtag of the memory allocation, which is \bench{0xa} in Fig.~\ref{fig:tripwire}, in \lastbits of the \shortgranule since \lastbyte (highlighted in red) of the \shortgranule is always unused. %
For {\granule}s other than the {\shortgranule}s, %
their {\memtag}s %
are randomly-generated and consistent with the pointer's \addrtag ({\eg} {\addrtag} and {\memtag} \bench{0xa} in Fig.~\ref{fig:tripwire}).
While resizing and freeing the allocation, \sys needs to propagate and clear out \lastbits of the \shortgranule.

\begin{figure}[t]
    \centering
    \small
    \includegraphics[width=0.9\linewidth]{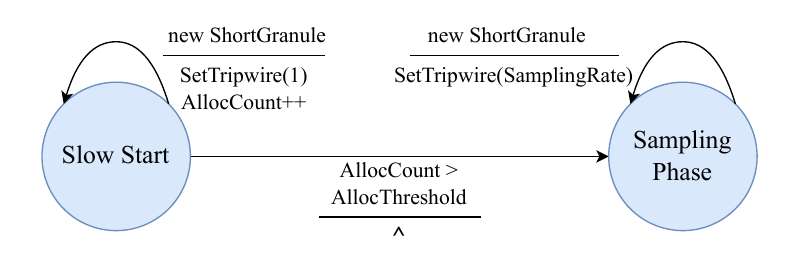}
    \caption{\small \sys's sampling-based \tripwire allocation.}
    \label{fig:tripwire-control}
\end{figure}

\pgheading{Sampling} %
Similar to \mte's probabilistic detection of {\violation}s, \sys also detects \intragranule \oob probabilistically via sampling, amortizing the overhead of additional software checks. 
As we show in Fig.~\ref{fig:tripwire-control}, while setting up {\tripwire}s, \sys employs sampling 
in two phases: \slowstart and \samplingphase.
In the \slowstart phase, for every new \shortgranule allocation (\bench{new ShortGranule}), \sys sets a \tripwire (\bench{SetTripwire(1)}),
while also incrementing the \alloccount. As \alloccount reaches above a threshold (\allocthreshold), \sys transitions from the \slowstart phase to the \samplingphase. In the \samplingphase, \sys %
sets the \tripwire only for the sampled ones (\bench{SetTripwire(SamplingRate)}). \sys uses a sampling algorithm similar to 
\gwpasan~\cite{serebryany2024gwp}.
\sys generates a random number (\bench{rand}) from \bench{[1, 2 * \samplingratenospace]}, and allocates a \tripwire after every \bench{rand} \shortgranule allocations. After setting up a \tripwire, \sys generates the random number \bench{rand} again for the subsequent \tripwire. %
By default, \sys sets both \samplingrate and \allocthreshold to 1000, which we empirically find to yield high bug detection capability (\julietoverflowsys, \S\ref{sec:eval-sensitivity}).
Both of these parameters are configurable.%

\revision{\pgheading{Implications of Sampling} For \sysns’s primary use case, \ie in-house testing, as stronger detection capability incurs higher overhead, \sys makes this tradeoff explicit with a dynamically configurable interface. By configuring this interface dynamically, \sysns’s probabilistic approach makes in-house testing techniques (\eg fuzzing) more effective. For example, as fuzzers observe interesting cases (\eg new path), they can increase \sysns’s bug detection capability. For uninteresting cases, fuzzers can instead prioritize performance. As fuzzing typically executes thousands of executions per second, many of which result in similar program behavior, the fuzzer is effective as long as it can detect the bug at one of these executions.}

\subsection{Byte-Granular Overflow Detection}
\label{sec:design-bounds-check}

\sys implements a byte-granular overflow detection algorithm (Algorithm~\ref{alg:bounds-checks}) in the \handler. %

\pgheading{Information Extraction}
In \mtesync, \sys gets necessary
information (\ie exception information) from the \fault including: the \exceptioninsn's precise location (\ie program counter, \bench{pc}), \faultaddr of the memory, values of general-purpose registers (\bench{x0}~\textasciitilde~\bench{x30}) and the stack pointer register (\bench{sp}). \sys uses this exception information to infer all inputs in Algorithm~\ref{alg:bounds-checks}. For \faultaddr (\bench{f}), \sys directly obtains it from the exception information. For the starting address (\bench{start}) and size (\bench{size}) of the memory access, \sys first reads the memory bytes corresponding to \bench{pc} and then decodes them to get the \exceptioninsn. 
Combining this decoded instruction with values of registers (\bench{x0}~\textasciitilde~\bench{x30}, \bench{sp}), \sys obtains \bench{start} and \bench{size}.
In ARM \mte, only memory access instructions, \ie ARM load and store instructions~\cite{loadstoreinsns},
can cause an \mte
\fault.
Based on the ISA specification of %
these instructions~\cite{loadstoreinsns}, \sys extracts the size of the memory access (\bench{size}) from the opcode, and decodes memory-access-related fields, \eg the base register index and the memory access offset, from the instruction encoding. \sys calculates the starting address of the memory access
(\bench{start}) with values of registers based on the addressing mode and memory-access-related fields of the instruction. \sys infers \memtag (\bench{memtag}) and \lastbits (\bench{metadata}) of the \shortgranule from the \faultaddr (\bench{f}). Assuming the \faultaddr is inside the \shortgranule, \sys uses the normal load instruction (\bench{LDRB}) to obtain \lastbits of the \shortgranule (\bench{metadata}), and uses \mte tag load instruction (\bench{LDG}) to obtain the \memtag of the \shortgranule (\bench{memtag}). \sys infers the \addrtag (\bench{addrtag}) with both the \exceptioninsn and values of registers. Similar to \bench{start} and \bench{size},
\sys decodes the \exceptioninsn to get the base register index and the offset. The offset can be either an immediate value in the instruction or a value in a register. If the offset is an immediate value, \sys takes \topbyte of the base register as the \addrtag (\bench{addrtag}). Otherwise, 
\sys calculates the \addrtag (\bench{addrtag}) with both \topbyte of the base register and \topbyte of the offset register.

\begin{algorithm}[t]
\caption{\small \sys's byte-granular overflow detection.} %
\label{alg:bounds-checks}
\begin{algorithmic}[1]
\small
\REQUIRE fault address (\bench{f}), starting address (\bench{start}) and size (\bench{size}) of the memory access, \addrtag (\bench{addrtag}),  \memtag (\bench{memtag}), and \lastbits (\bench{metadata}) of the \shortgranule. 
\ENSURE \TRUE~if the memory access is benign, \FALSE~if it is a \violation.
\IF{$\bench{memtag} = 0$ \OR $\bench{addrtag} = 0$}
    \RETURN \FALSE
\ENDIF

\IF{$\bench{addrtag} \neq \bench{metadata}$}
    \RETURN \FALSE
\ELSE
    \STATE $\bench{shortgranule} \gets \bench{f} \mathbin{\&} \lnot(16 - 1)$ \COMMENT starting address of the \shortgranule
    \STATE $\bench{permitted} \gets \bench{shortgranule} + \bench{memtag}$ \COMMENT end address of addressable bytes in the \shortgranule
    
    \STATE $\bench{attempted} \gets \bench{start} + \bench{size}$ \COMMENT end address of the memory access
    \IF{$\bench{attempted} \le \bench{permitted}$}
        \RETURN \TRUE
    \ELSE
        \RETURN \FALSE
    \ENDIF
\ENDIF
\end{algorithmic}
\end{algorithm}

\pgheading{Overflow Detection} \sys determines whether a memory access is benign or a \violation based on a set of addressability properties. In \sys, a byte's addressability on the heap follows the following properties:

\textbf{(P1)}: The byte is non-addressable if its \memtag is 0.

\textbf{(P2)}: The byte is non-addressable for %
a pointer with \addrtag 0.

\textbf{(P3)}: The byte is addressable when its \memtag matches the \addrtag of the pointer.

\textbf{(P4)}: The byte is addressable when it is the addressable part of a \shortgranule, and \lastbits of the \shortgranule match the \addrtag of the pointer.

As we show in Algorithm~\ref{alg:bounds-checks}, \sys rejects a memory access if
its \addrtag or \memtag is zero based on P1 and P2 (lines 1\textasciitilde 2). As ARM \mte protects memory regions by tagging them with a randomly selected \memtag which is always not zero~\cite{mteefficient}, a \granule with \memtag being zero means that it is either not allocated or not protected by \mte. Therefore, P1 and P2 prevent \oob that are across heap, stack, or code segments.
A \fault occurs when P3 is violated. \sys implements Algorithm~\ref{alg:bounds-checks} in the \handler to analyze whether the memory access that causes the \fault is benign or a \violation based on P4. It rejects the memory access if \lastbits of the \granule is not the same as the \addrtag of the pointer (line 3\textasciitilde 4), which indicates that the \fault is not caused by a \tripwire.

\begin{figure}[t]
    \centering
    \includegraphics[width=\linewidth]{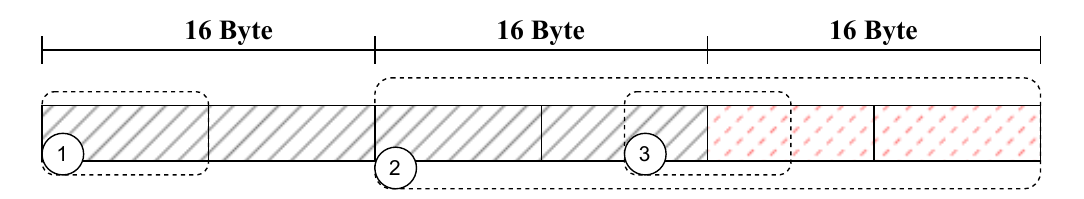}
    \caption{\small Examples of memory access patterns of ARM load and store instructions. The red block represents the tripwire. The memory access can be unaligned, \eg access 3, and may access multiple {\granule}s, \eg access 2 and 3.
    }
    \label{fig:loadstore}
\end{figure}

As we show in Fig.~\ref{fig:loadstore}, ARM load and store instructions can be unaligned, accessing multiple {\granule}s. For example, a store %
instruction (access 2) can write two 16-byte values from two %
registers into two {\granule}s in the memory. Similarly, a load %
instruction (access 3) %
can read memory starting from any address, %
which may cross the \granule's boundary.
Therefore, \sys determines whether %
the instruction accesses \shortgranule's addressable bytes
by calculating both the end address of addressable bytes in the \shortgranule (\bench{permitted}, line 6\textasciitilde 7) and the end address of the bytes the instruction accesses (\bench{attempted}, line 8). As we show in Algorithm~\ref{alg:bounds-checks}, \sys only permits memory accesses if the \bench{attempted} last byte of the memory access is not larger than the \bench{permitted} last byte of the \shortgranule (line 9\textasciitilde 12). %
As long as the instruction is accessing at least one of the non-addressable bytes in the \shortgranule,
the end address of the memory access (\bench{attempted})
is always going to be bigger than 
the end address of addressable bytes in the \shortgranule (\bench{permitted}). 
Consequently, \sys will flag such an access as a \violation.

\subsection{Tag Mismatch Recovery}
\label{sec:design-recovery}

In the \handler, \sys resumes the program execution if %
the memory access is benign. However, resuming the program execution after a \fault is not trivial. 
If \sys resumes the program without removing the \tripwire, %
the program will access the \tripwire, causing a \fault again.
On the other hand, removing the \tripwire before resuming the program avoids the \fault but disables the byte-granular \boundscheck in the \granule ever after. 
\sys addresses this issue by presenting a \textit{delegation-escalation-revocation} approach. %
Overall, \sys achieves the following goals:

\pgheading{(G1)} The processor executes all instructions in the program order, no matter whether a \fault occurs or not.

\pgheading{(G2)} \sys revokes the permission as soon as possible. %

\pgheading{(G3)} After \sys revokes the permission, accessing the \tripwire will trigger a \fault.

\begin{figure}[t]
    \centering
    \includegraphics[width=\linewidth]{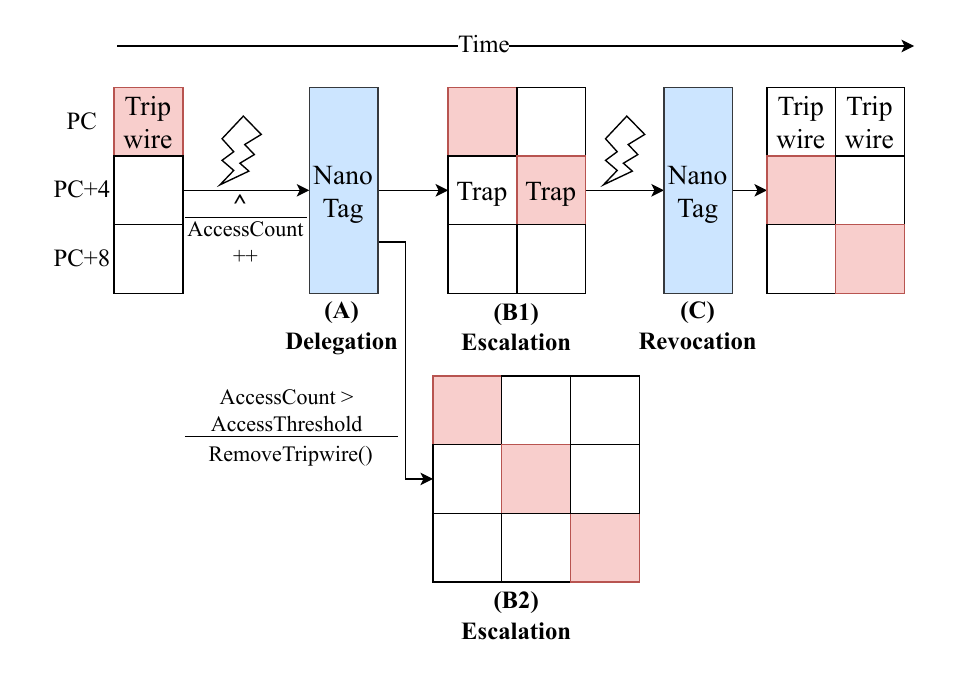}
    \caption{\small Delegation-escalation-revocation in \sys. A red block denotes the program instruction processor currently executes, blue blocks represent \sys's byte-granular overflow detection and tag mismatch recovery, and lightning symbols denote \fault and \trap. 
    }%
    \label{fig:tag-mismatch-recovery}
\end{figure}

\pgheading{Delegation} \sys resolves the \fault by manipulating the tags involved in the memory access. %
In particular, as we show in Fig.~\ref{fig:tag-mismatch-recovery}(A), \sys %
removes the \tripwire by
setting the \tripwiretag to be the same as the \addrtag of the pointer. %
As \sys needs to reset the \tripwiretag during \emph{revocation}, %
\sys swaps the \memtag and \lastbits of the corresponding \shortgranule. %
The program behavior stays the same as if the \fault had not occurred, since the program executes the same instruction (G1).
\sys does not modify the \addrtag of the pointer, since
a single %
instruction can access multiple {\granule}s that do not share the same \memtag, as we show for accesses 2 and 3 in Fig.~\ref{fig:loadstore}. Modifying the \addrtag of the pointer to match one of the {\memtag}s may %
cause a \fault in another \granule.
\sys also avoids emulating the \exceptioninsn %
as, %
in software, it is challenging to precisely emulate %
instructions that require atomicity (\eg \bench{RCWCAS}) or exclusive access (\eg \bench{STTXR}), %
causing the program to have a different behavior. %

\pgheading{Escalation} After \emph{delegation}, the program %
can use the pointer to access every byte in the delegated \shortgranule.
If \sys does not revoke the permission promptly, %
the program will run in an escalated state for a long time, violating (G2). 
Therefore, \sys sets a \textit{\trap} right after the \exceptioninsn to revoke the permission, as we show in Fig.~\ref{fig:tag-mismatch-recovery}(B1).
A \trap is an instruction that causes an exception
during execution,
such as %
instructions ARM hardware breakpoints~\cite{breakpoint, debugregs} observe,
software \breakpoint instructions (\eg \bench{BRK}), or illegal instructions. 
After setting the \trap, \sys resumes the program execution from the \exceptioninsn. \sys's escalation has limitations (\eg %
\exceptioninsn at the end of a function) %
that we describe in \S\ref{sec:discussion}. %

\pgheading{Revocation} 
Once the program's execution resumes, the \exceptioninsn accesses the memory without causing a \fault. 
The program then executes the next instruction that contains a \trap.
This \trap causes an exception, which \sys intercepts to revoke the permission for the delegated \shortgranule.
In particular, \sys sets the \memtag of the \shortgranule back to the original \tripwiretag, by swapping the \memtag and \lastbits of the corresponding \shortgranule again, as we show in Fig.~\ref{fig:tag-mismatch-recovery}(C).
After \emph{revocation},
accessing the \shortgranule will again trigger a \fault (G3).
\sys also removes the \trap during \emph{revocation}, resuming the program from the current instruction.

\pgheading{\tripwirecap Access Control} 
As \sys %
recovers after a \fault, accessing the same \tripwire repeatedly may introduce significant performance slowdown. \sys avoids the potential overhead %
by adopting an access control mechanism as we show in Fig.~\ref{fig:tag-mismatch-recovery}.
In particular, \sys maintains a counter, \acccount, in the padded bytes of the \shortgranule,
varying from 1 to 15 bytes.
In case of a single-byte padding, \sys leverages the most significant 4 bits of the padding for \acccount, as the padding's least significant 4 bits already contain the valid \addrtag.
In case of a multi-byte padding, \sys could utilize at least 12 bits for \acccount. %
In both cases, \sys increments a \shortgranule's \acccount every time the corresponding \tripwire causes a \fault. 
When a \shortgranule's \acccount reaches the largest number the counter could hold (\eg 15 for a 4-bit counter), \sys removes the \shortgranule's \tripwire. %
\sys supports configuring another parameter, \accthreshold. As a \shortgranule's \acccount reaches \accthreshold, \sys also removes the %
\tripwire, as we show in Fig.~\ref{fig:tag-mismatch-recovery}(B2). %
By default, \sys sets \accthreshold to 64, which we empirically find to provide high bug detection capability (\julietoverflowsys, \S\ref{sec:eval-sensitivity}).

\subsection{\sys Bug Report}
\label{sec:design-bug-report}

Once \sys detects a \violation, it provides a detailed report to help developers pinpoint the bug's root cause.
\sys reports program's \bench{pc}, \faultaddr, values of registers, \addrtag, and \memtag.
If the \violation is an \intragranule \oobsingle, \sys also reports the number of addressable bytes in the \shortgranule, while also listing the number of bytes the \exceptioninsn accesses in the \shortgranule to cause the bug.
\revision{To reliably reproduce bugs, \sys supports increasing its bug detection capability by adjusting its configurable parameters, \eg by setting \allocthreshold as \bench{INT\_MAX}, assuming no tag collision for \mte.}

\section{Implementation}
\label{sec:impl}

\begin{table*}[t]
    \centering
    \caption{\small {\violationcap}s detected in \juliet. \sys detects a similar number (\julietoverflowsys) of heap-based buffer overflow bugs as ASAN (\julietoverflowasan).
    }
    \label{tab:juliet-bad-eval}
    {
    \footnotesize
    \begin{tabular}{c|c|c|c|c|c|c}
    \hline
        \textbf{CWE ID} & \textbf{Total} & \textbf{ASAN} & \textbf{\makecell[c]{\gwpasan}}& \textbf{\makecell[c]{Scudo (ASYNC)*}} & \textbf{\makecell[c]{Scudo (SYNC)*}} & \textbf{\sys}\\
    \hline
        \makecell[c]{\bench{\footnotesize CWE122}\\(heap-based buffer overflow)} & 3438 & 98.66\% & \makecell[c]{23.15\% \\± 0.24\%} & \makecell[c]{75.60\% \\± 0.13\%} & \makecell[c]{75.68\% \\± 0.10\%} & \makecell[c]{97.57\% \\± 0.15\%} \\\hline
        \makecell[c]{\bench{\footnotesize CWE415}\\(double free)}                & 818  & 100\% & \makecell[c]{98.60\% \\± 0.23\%} & \makecell[c]{98.43\% \\± 0.14\%} & \makecell[c]{98.45\% \\± 0.19\%} & \makecell[c]{98.37\% \\± 0.12\%} \\\hline
        \makecell[c]{\bench{\footnotesize CWE416}\\(use after free)}             & 393  & 100\% & \makecell[c]{0\%} & \makecell[c]{96.54 \\± 0.60\%} & \makecell[c]{96.94\% \\± 0.58\%} & \makecell[c]{96.69\% \\± 0.44\%} \\
    \hline
    \end{tabular}
    }\\
    *The variance comes from \mte's tag collision across multiple runs.
\end{table*}

We implement \sys using a standalone user-mode signal handler~\cite{signalhandler} and \scudofull~\cite{androidscudo}. In particular, \sys's signal handler consists of \handlerloc lines of code. We implement \sys's configurable parameters using environment variables~\cite{envvar}. We open-source \sys at \url{https://github.com/ice-rlab/NanoTag}.

\pgheading{Sampling-Based \tripwirecap Allocation} We implement \sys's sampling-based \tripwire allocation by modifying \scudomodifyloc lines of code in \scudo.
We implement \sys's sampling-based \tripwire allocation by modifying \scudo's primary allocator.
We compile both \scudo and \sys with the \bench{O2} optimization level.

\pgheading{Overflow Detection} \mte raises a \fault as a segmentation fault, \ie a \bench{SIGSEGV} signal with codes 8 and 9 for \mte ASYNC and SYNC mode, respectively. Consequently, 
we implement \sys's \handler as a user-space signal handler using \bench{sigaction} with \bench{SA\_SIGINFO} and \bench{SA\_RESTART} flags. %
Our prototype supports a variety of load and store instructions, including regular, atomic, and SIMD instructions.
We find that some functions in 
\bench{glibc}~\cite{glibc} introduce out-of-bound read accesses in their assembly implementations. For example, the \bench{strcpy}'s ARM assembly implementation~\cite{strcpy} uses SIMD load instructions to iteratively read from the source string and then checks if it has reached the null terminator. While there exist prior ARM efforts~\cite{compatiblestrcpy} to make \bench{strcpy} compatible with \mte, it will still require additional engineering efforts to make \bench{glibc} compatible with byte-granular \boundscheck. Consequently, to avoid rewriting \bench{glibc}, \sys's \handler skips the {\fault}s these %
instructions cause.

\pgheading{Tag Mismatch Recovery} We implement \sys's tag mismatch recovery mechanism by setting up ARM \bench{BRK} instructions, as we could not set hardware breakpoints on \device. When the processor executes this \bench{BRK} instruction, it generates a \bench{SIGTRAP} signal, %
which \sys catches with another signal handler.
To differentiate from other \bench{SIGTRAP} signals, \sys uses a special immediate value while setting the \bench{BRK} instruction.

\pgheading{Bug Report} %
\sys's signal handler parses the \bench{ucontext} argument~\cite{ucontext}, extracting information, \eg \bench{pc}, \faultaddr, and values of registers, to generate the bug report, similar to what we describe in \S\ref{sec:design-bounds-check}.

\section{Evaluation}
\label{sec:eval}

In this section, we experimentally evaluate \sys to answer the following research questions:

\pgheading{RQ1} How effectively does \sys detect {\violation}s at byte granularity?

\pgheading{RQ2} How efficiently does \sys detect {\violation}s at byte granularity in terms of run-time overhead? %

\pgheading{RQ3} How does \sys generalize to real-world applications in terms of run-time efficiency and fuzzing throughput? %

\pgheading{RQ4} How do \sys's parameters affect its bug detection capability and run-time overhead?

\subsection{Experimental Methodology}
\label{sec:eval-setup}

Similar to prior work~\cite{stickytags, momeu2025iubik}, we perform
all experiments
on a rooted \devicemodel with \termux~\cite{termux}. We first describe our hardware and software setup. Then, we provide a brief overview of our benchmarks and performance metrics.

\pgheading{Hardware} The \devicemodel is one of the first handsets with \mte support~\cite{projectzero}. This device has four ``big'' Cortex-A715 cores, 
four ``little'' Cortex-A510 cores,
and one ``prime'' Cortex-X3 processor core, with a frequency range of 402-2367 MHz, 324-1704 MHz, and 500-2914 MHz, respectively.
The device has 12 GB LPDDR5 RAM. 

\pgheading{Software} The \devicemodel uses the Android %
15.
On this system, we first install \termux~\cite{termux}, an \android terminal emulator that provides a Linux development environment on \android OS. As \termux launches programs directly by calling the \bench{execve()} system call, it incurs minimal overhead compared to a virtualized environment~\cite{mantermux}. We also leverage \bench{chroot}~\cite{chroot}, a Linux command, to change the system's root directory to a specific path, the Ubuntu 22.04 root file system (\bench{rootfs}), to emulate a Linux development environment. In particular, we use the Linux kernel version of %
5.15.148. 
As compiler, %
we use LLVM's front-end compiler clang %
14~\cite{clang14}.

\pgheading{Benchmarks} We evaluate \sys's bug detection capability (\S\ref{sec:eval-security}) and performance efficiency (\S\ref{sec:eval-performance}) using \juliet~\cite{juliet} and \spec~\cite{specbenchmark}, respectively. 
\revision{We evaluate \sys's generality to real-world applications using \geekbench~\cite{geekbench}, \bench{Memcached}~\cite{memcached}, \bench{LevelDB}~\cite{leveldb}, \bench{RocksDB}~\cite{rocksdb}, and \magma~\cite{hazimeh2020magma} (\S\ref{sec:eval-case-study}).}

\pgheading{Metrics} For performance efficiency, we report overhead numbers as the percentage increase in benchmark execution time with a sanitized configuration (\eg ASAN, \sys, etc.) in comparison to the baseline configuration (\ie \mte-disabled \scudo). We run each benchmark multiple times to report both arithmetic and geometric means.

\subsection{Bug Detection Capability}
\label{sec:eval-security}

We study \sys's {\violation} detection capability using \juliet~\cite{juliet},
comparing the results with both \scudo in \mte SYNC and ASYNC modes, \gwpasan, and ASAN.

\pgheading{Settings} While using \juliet to evaluate \sys, we use the same experimental setting as our analysis (\S\ref{sec:analysis}). 
For example, we evaluate \sys for both benign and buggy versions of each CWEs. 
We run \juliet with \scudo in MTE SYNC and ASYNC modes, \gwpasan, ASAN, and \sys. 
We compile all test cases with \bench{O0} so that compiler can not optimize them to avoid {\violation}s. Due to \mte's probabilistic nature, for \scudo and \sys, we run \juliet for ten iterations and report the average number. We run \juliet directly on \android 15 using \termux.

\pgheading{Results} 
As we discussed in \S\ref{sec:analysis-security-measurement}, ASAN, \gwpasan, and \scudo do not label any of the benign test cases in the \juliet as {\violation}s. Similarly, \sys flags none of the benign test cases as a \violation in any iteration of the experiment.
Therefore, we only show the results for the buggy test cases of \juliet in Table~\ref{tab:juliet-bad-eval}. In particular, Table~\ref{tab:juliet-bad-eval} shows the percentage of bugs ASAN, \gwpasan, \scudo, and \sys detect for different CWEs in \juliet. 
As we also show in \S\ref{sec:analysis-security-measurement}, ASAN detects \julietoverflowscudolessthanasan of more heap-based buffer overflow bugs (\bench{CWE122}) than \scudo in \mte SYNC or ASYNC mode, primarily due to \intragranule \oob.
In contrast, \sys identifies these \intragranule \oob successfully, detecting \julietoverflowsys of all heap-based buffer overflow bugs, similar to \julietoverflowasan of bugs ASAN detects.
The \julietoverflowsyslessthanasan gap between \sys and ASAN in \bench{CWE122} is primarily due to \mte's probabilistic nature.
As we mention in \S\ref{sec:intro} and \S\ref{sec:analysis}, \mte uses only 4-bit tags due to constraints on tag storage and TBI.
Consequently, two random tags may collide with each other with a probability of 6.25\% ($\frac{1}{16}$ as $2^4=16$). Prior work~\cite{unterguggenberger2023multitag} aims to avoid such tag collisions by utilizing a multi-granular memory tagging in hardware.
For double free (\bench{CWE415}) and \uaf (\bench{CWE416}), \sys performs similarly to \scudo, detecting \julietdoublefreesys and \julietuafsys double free and \uaf bugs, respectively. 
Similar to heap-based buffer overflow bugs, \mte's probabilistic nature prevents \sys from detecting the rest of the %
temporal bugs.

\takeaway{Detecting \intragranule \oob, \sys performs similar %
to ASAN 
in \juliet.}

\subsection{Run-Time Overhead}
\label{sec:eval-performance}

\begin{figure}[t]
    \centering
    \includegraphics[width=\linewidth]{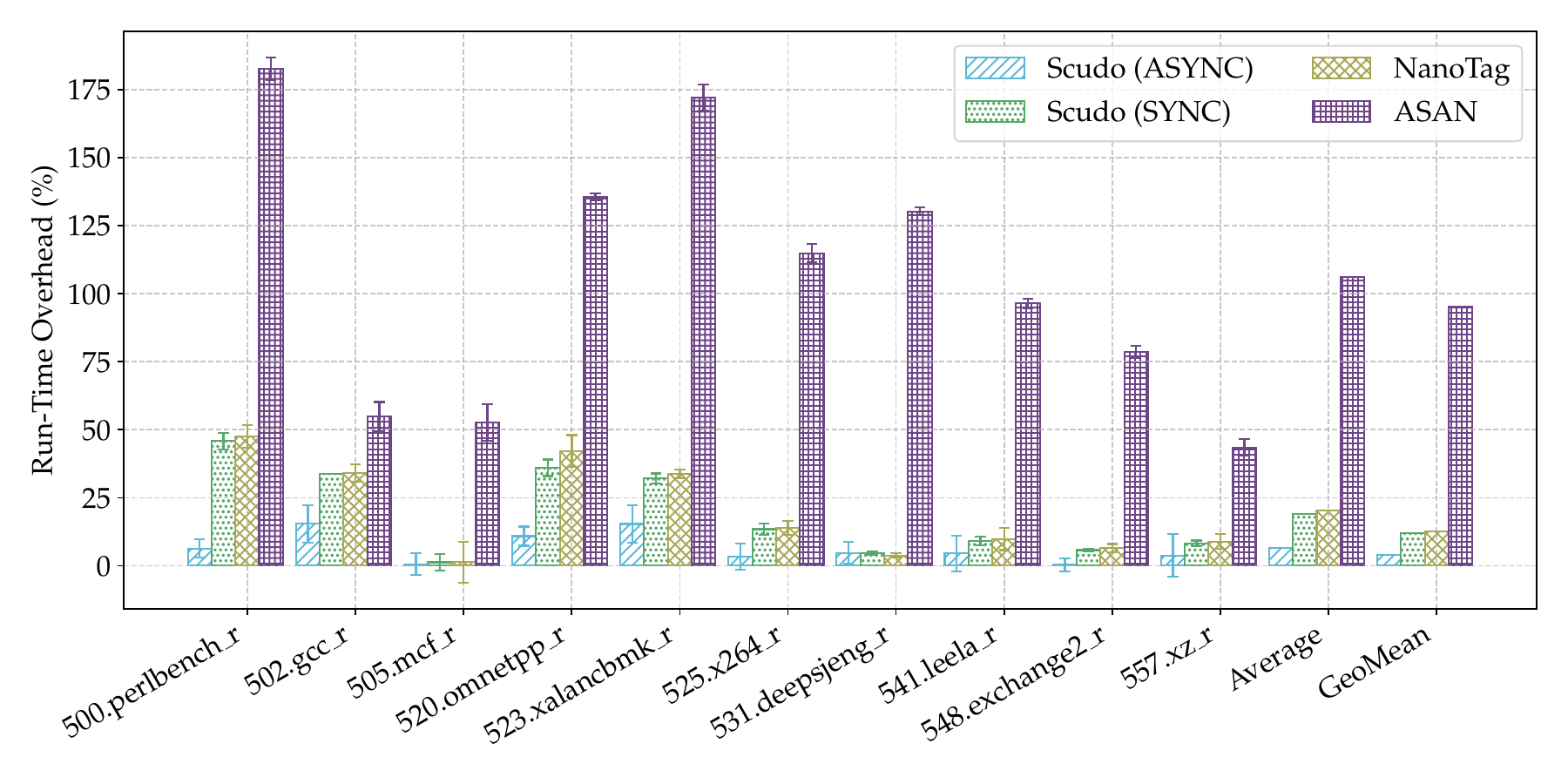}
    \caption{\small The run-time overhead of \mte-enabled \scudo, \sys, and ASAN for \spec compared to a \scudo baseline with \mte disabled. In terms of geometric mean, \sys incurs an overhead of \specintsys similar to \scudo's \specintsync overhead in \mtesync.}
    \label{fig:spec-slowdown-eval}
\end{figure}

We study \sys's run-time overhead using \spec~\cite{specbenchmark}. 
Specifically, we use \specint to measure the performance overhead \sys, \mte-enabled \scudo, and ASAN incur over a \scudo baseline with \mte disabled.
We show the results in Fig.~\ref{fig:spec-slowdown-eval}. Reusing padding bytes to store metadata, \sys avoids incurring any memory overhead.

\pgheading{Settings} Using \spec version (1.1.9), we evaluate \sys for all \numbenchmarks \specint, including the ones written in Fortran. 
We only use \specint similar to prior work~\cite{stickytags} as \device's 12 GB physical memory is not sufficient to run SPECspeed Integer benchmarks that require 16 GB of physical memory.
We cross-compile all benchmarks on an ARM server with Cavium ThunderX2 CPUs~\cite{thunderx2} provided by Chameleon Cloud~\cite{chameleon}, and run them on \device using \bench{chroot} Ubuntu 22.04. While compiling all benchmarks with \bench{clang}, we use the \bench{O3} optimization level.
We use this same set of binaries to evaluate \scudo and \sys via \bench{LD\_PRELOAD}, following the SPEC tutorial~\cite{avoidruncpu}. As ASAN requires additional compilation flags, we evaluate ASAN using a separate set of binaries compiled with the required flags (\bench{-fsanitize=address -fsanitize-recover=address}).

\pgheading{Results} Fig.~\ref{fig:spec-slowdown-eval} shows the run-time overhead of \scudo, \sys, and ASAN for \spec. %
In terms of geometric mean, \sys incurs a performance overhead of \specintsys, significantly outperforming ASAN (\specintasan). Even in the worst case, \sys slows down \bench{500.perlbench\_r} by \perlsys, which is only around $\frac{1}{4}$ of ASAN's \perlasan overhead. 
Compared to the \specintsync overhead of \scudo in \mtesync, 
\sys incurs only an additional \specintsysmorethansync overhead. 
Its largest gap to \scudo in \mtesync occurs in \bench{520.omnetpp\_r}, which is \specintsysmorethansyncmax.
For \scudo, we also observe a noticeable gap between MTE SYNC and ASYNC modes. In terms of geometric mean, \scudo only incurs a run-time overhead of \specintasync in \mteasync, which is \specintsyncmorethanasync lower than the overhead of \scudo in \mtesync. The largest gap between MTE SYNC and ASYNC modes also occurs for \bench{500.perlbench\_r}, which is \perlsyncmorethanasync. 
Reducing \mtesync's overhead will also make \sys more efficient, which we plan as future work.

\takeaway{In terms of geometric mean, \sys incurs a run-time overhead of \specintsys for \spec.}

\subsection{Real-World Case Studies}
\label{sec:eval-case-study}

\begin{table}[t]
    \centering
    \caption{\small Run-time overhead of \scudo with \mte ASYNC and SYNC modes, \sys, and \valgrind~\cite{nethercote2007valgrind} on \geekbench~\cite{geekbench} compared to the \scudo baseline with \mte disabled. On the closed-source \geekbench benchmarks, \sys adds only \geekbenchsysoversync overhead over the \mte SYNC mode.}
    \label{tab:case-study}
    {
    \footnotesize%
    \begin{tabular}{c|c|c|c|c}
    \hline
         & \textbf{\makecell[c]{Scudo\\ (ASYNC)}} & \textbf{\makecell[c]{Scudo\\ (SYNC)}} & \textbf{\sys} & \textbf{\makecell[c]{\valgrind\\ Memcheck}}\\
    \hline%
        \makecell[c]{\geekbench} & 1.96\% & 3.76\% & 4.99\% & 1348.60\% \\
    \hline
    \end{tabular}
    }
\end{table}

\subsubsection{Closed-Source Application} We study \sys's run-time overhead on real-world closed-source applications using \geekbench~\cite{geekbench}, one of the most popular benchmarking apps with 1 million downloads on Google Play Store. %

\pgheading{Settings} We use the \geekbench Linux/AArch64 Preview version~\cite{previewgeekbench}, which runs the same set of benchmarks as the \android or \ios version of \geekbench~\cite{playstoregeekbench, appstoregeekbench}. The closed-source binary consists of 16 benchmarks that always run sequentially, one benchmark after another. 
As the binary does not provide an option to run a single benchmark separately, while evaluating \sys, we set \allocthreshold to 100,000, much higher than its default value (1000), so that we can measure \sys's overhead for all benchmarks.
Similar to \spec, we run \geekbench on chroot Ubuntu 22.04 with a single core and measure the absolute run time. 
For baseline, we use \scudo with \mte disabled and compare its performance against \sys and \scudo with \mte ASYNC and SYNC modes.
\retrowrite~\cite{dinesh2020retrowrite}, the \sotaadj static binary instrumentation tool to sanitize closed-source binaries, does not support the C++ exception handling mechanism \geekbench requires.
Consequently, for \geekbench, we compare \sys's performance against %
\valgrind~\cite{nethercote2007valgrind}'s Memcheck tool~\cite{seward2005using}. %

\pgheading{Results} As we show in Table~\ref{tab:case-study},
\sys incurs a run-time overhead of only \geekbenchsys on \geekbench, which is close to the overheads \scudo ASYNC (\geekbenchasync) and SYNC (\geekbenchsync) modes incur, and much lower than the overhead of \valgrind (\geekbenchvalgrind). 
For all configurations, Table~\ref{tab:case-study} shows the overheads of the first 14 out of 16 benchmarks of \geekbench. While running the 15th benchmark (\bench{ray-tracer}) with \sys, we observe an unexpected segmentation fault. The root cause of the segmentation fault is a software breakpoint (\bench{BRK} instruction) \sys sets up during \emph{escalation} to restore a \tripwire. For some specific instruction addresses of \bench{ray-tracer}, software breakpoints cause a segmentation fault instead of causing an exception. While testing the benchmark with \bench{GDB} manually, we also could not set up breakpoints for those addresses. In the future, we will explore setting up ARM hardware breakpoints~\cite{breakpoint, debugregs}, currently not supported on \device, to solve this problem.

\begin{table}[t]
\centering
\revision{
\caption{\small Large real-world applications, their versions, benchmarks, and workloads we study.}
\footnotesize%
\label{tab:db-settings}
\begin{tabular}{c|c|c|c}
\hline
\textbf{Applications} & \textbf{Versions} & \textbf{Benchmarks} & \textbf{Workloads} \\ \hline
\bench{\footnotesize Memcached} & 1.6.14 & \bench{\footnotesize mc-benchmark}~\cite{mc-benchmark} & \bench{\footnotesize GET}, \bench{\footnotesize SET} \\ \hline
& & & \multirow{1}{*}{\makecell[c]{\bench{\footnotesize fillseq},\\ \bench{\footnotesize readseq},\\ \bench{\footnotesize fillrandom},\\ \bench{\footnotesize readrandom},\\ and\\ \bench{\footnotesize readreverse}}} \\
\bench{\footnotesize LevelDB} & 1.22 & \bench{\footnotesize db\_bench}~\cite{ldb-dbbench} & \\ 
& & & \\ \cline{1-3}
& & & \\
\bench{\footnotesize RocksDB} & 10.10.1 & \bench{\footnotesize db\_bench}~\cite{rdb-dbbench} & \\
& & & \\
\hline
\end{tabular}
}
\end{table}

\begin{table}[t]
    \centering
    \revision{
    \caption{\small Run-time overhead of \sys and ASAN on three large real-world applications. %
    }
    \label{tab:real-world-app}
    {
    \footnotesize
    \begin{tabular}{c|c|c|c}
    \hline
        \multirow{2}{*}{\textbf{Target}} & \multicolumn{2}{c|}{\textbf{\sys Slowdown}} & \textbf{ASAN} \\ \cline{2-3}
         & \textbf{\samplingratenospace=10} & \textbf{\samplingratenospace=100} & \textbf{Slowdown} \\ 
    \hline
        \bench{\footnotesize Memcached} & 6.07\% & 1.30\% & 22.53\% \\
        \bench{\footnotesize LevelDB}   & 16.93\% & 12.56\% & 95.26\% \\
        \bench{\footnotesize RocksDB}   & 18.35\% & 17.75\% & 1698.31\% \\ \hline
        GeoMean   & 12.35\% & 6.13\% & 153.90\% \\
    \hline
    \end{tabular}
    }
    }
\end{table}

\begin{table}[t]
    \centering
    \caption{\small Fuzzing throughput slowdown for \sys and ASAN on \magma benchmarks compared to the \scudo baseline. 
    \sys slows down fuzzing throughput by \fuzzingsys, $\frac{1}{7}$ of ASAN's slowdown.
    }
    \label{tab:fuzzing}
    {
    \footnotesize%
    \begin{tabular}{c|c|c|c|c}
    \hline
        \textbf{Target} & \textbf{\makecell[c]{\bench{\footnotesize.text} Size \\(Baseline)}} & \textbf{\makecell[c]{Baseline \\(exec/s)}} & \textbf{\makecell[c]{\sys \\Slowdown}} & \textbf{\makecell[c]{ASAN \\Slowdown}}\\
    \hline
        \makecell[c]{libpng} & 277.3 KB & 28516.9 & 17.83\% & 35.70\% \\
        \makecell[c]{libxml2} & 1.1 MB & 9243.23 & 7.26\% & 113.49\% \\
        \makecell[c]{poppler} & 4.4 MB & 356.57 & 30.85\% & 339.32\% \\\hline
        \makecell[c]{GeoMean} &  &  & 15.86\% & 111.20\% \\
    \hline
    \end{tabular}
    }
\end{table}

\begin{table*}[t]
    \centering
    \caption{\small %
    Bug detection capability (\%) and run-time overhead (\%) for \sys's different configurations.}
    \label{tab:sensitivity}
    \footnotesize
    \begin{tabular}{ccc|ccc|ccc}
        \toprule
        \multicolumn{3}{c|}{\textbf{(a) \accthreshold}} & \multicolumn{3}{c|}{\textbf{(b) \allocthreshold}} & \multicolumn{3}{c}{\textbf{(c) \samplingrate}} \\
        \cmidrule(r){1-3} \cmidrule(lr){4-6} \cmidrule(l){7-9}
        \textbf{Value} & \textbf{\makecell[c]{Bug Detection\\Capability (\%)}} & \textbf{\makecell[c]{Run-Time\\Overhead (\%)}} & \textbf{Value} & \textbf{\makecell[c]{Bug Detection\\Capability (\%)}} & \textbf{\makecell[c]{Run-Time\\Overhead (\%)}} & \textbf{Value} & \textbf{\makecell[c]{Bug Detection\\Capability (\%)}} & \textbf{\makecell[c]{Run-Time\\Overhead (\%)}} \\
        \midrule
        32    & 97.38 & 47.08 & 100   & 97.49 & 47.36 & 10    & 97.52 & 100.95 \\
        64    & 97.49 & 47.60 & 1000  & 97.41 & 47.60 & 100   & 97.20 & 61.23  \\
        128   & 97.50 & 59.00 & 10000 & 97.45 & 58.02 & 1000  & 97.44 & 47.60  \\
        256   & 97.32 & 59.32 &       &       &       &       &       &        \\
        \bottomrule
    \end{tabular}
\end{table*}

\revision{
\subsubsection{Large Real-World Applications} We evaluate \sys's run-time overhead on three large real-world applications: \bench{Memcached}~\cite{memcached}, \bench{LevelDB}~\cite{leveldb}, and \bench{RocksDB}~\cite{rocksdb}.

\pgheading{Settings} We set up real-world applications similar to prior work~\cite{ugur-one-profile-fits-all-osr-2022}. For these applications, we summarize their versions, benchmarks, and workloads in Table~\ref{tab:db-settings}.
As the current prototype of \sys's slow start phase supports only a limited number of system calls, we disable this phase for these system-call-heavy applications~\cite{ugur-one-profile-fits-all-osr-2022}. As we only evaluate \sys's sampling phase for these applications, we set the sampling rate (\samplingrate) to 10 and 100 instead of 1000 to compensate for the lack of the slow start phase. For both sampling rates, we compare \sys against ASAN in terms of run-time overhead over the baseline \scudo with \mte disabled.

\pgheading{Results} As we show in Table~\ref{tab:real-world-app}, \sys incurs a geometric mean run-time overhead of up to \appsys on these three large real-world applications even with a \samplingrate of 10. %
For a \samplingrate of 100, \sys slows down these applications by only \appsyshs.
In contrast, ASAN incurs a run-time overhead of \appasan for these applications. %
}

\subsubsection{Fuzzing Throughput}
We evaluate %
\sys's generality on fuzzing 
using the \magma fuzzing benchmark~\cite{hazimeh2020magma}.

\pgheading{Settings} Using the %
\magma version 1.2.1~\cite{magmagithub}, we fuzz the \bench{libpng}, \bench{libxml2}, and \bench{poppler} targets with \aflpp~\cite{fioraldi2020afl++} in the persistent mode~\cite{persistentmode} on \device, in the absence of accessible \mte-enabled servers~\cite{ampereone,kaushik2025optimized}. Following the prior work~\cite{xu2017designing}, we run fuzzers for 5 minutes to report the average throughput of 5 %
fuzzing campaigns. We measure %
slowdowns
\sys and ASAN incur over the \scudo
baseline.%

\pgheading{Results} Table~\ref{tab:fuzzing} shows the fuzzing throughput slowdowns of \sys and ASAN. %
\bench{libpng}, \bench{libxml2}, and \bench{poppler} represent diverse fuzzing targets, from small to larger sizes, as we show the \bench{.text} section size and baseline fuzzing throughput in Table~\ref{tab:fuzzing}. In terms of geometric mean, \sys slows down the fuzzing throughput by only \fuzzingsys, which is around $\frac{1}{7}$ of ASAN's slowdown (\fuzzingasan). 
\sys's highest slowdown (30.85\%) occurs in \bench{poppler}, which is %
$\frac{1}{10}$ of ASAN's slowdown.%

\takeaway{\revision{
\sys incurs a negligible run-time overhead of only \geekbenchsys on \geekbench, a real-world closed-source application,
incurs up to \appsys run-time overhead on three large real-world applications: \bench{Memcached}, \bench{LevelDB}, and \bench{RocksDB}, %
and slows down fuzzing throughput by only \fuzzingsys on \magma.
}
}

\subsection{Sensitivity Study}
\label{sec:eval-sensitivity}

As we describe in \S\ref{sec:design}, \sys includes three configurable parameters: \accthreshold, \allocthreshold, and \samplingrate. We now study how these parameters affect \sys's performance overhead and bug detection capabilities. As we vary the values of these parameters, we investigate \sys's performance overhead using the \bench{500.perlbench\_r} benchmark, which suffers the highest run-time overhead ($47.6\%$) among \spec, while studying \sys's bug detection capabilities using \bench{CWE122} (heap-based buffer overflow) in \juliet.

\pgheading{Settings} The default values of \accthreshold, \allocthreshold, and \samplingrate are 64, 1000, and 1000, respectively. While studying the sensitivity of one parameter, we use the default values for the remaining parameters. %

\pgheading{Results} We show the results %
in 
Table~\ref{tab:sensitivity}. %
In Table~\ref{tab:sensitivity}~(a), as we increase \accthreshold from 32 to 256, \sys's run-time overhead increases from 47\% to 59\%, without affecting its bug detection capability that remains around 97\%.
Similarly, as we increase \allocthreshold from 100 to 10,000 in Table~\ref{tab:sensitivity}~(b), and decrease \samplingrate from 1000 to 10 in Table~\ref{tab:sensitivity}~(c), \sys's run-time overhead increases from 47\% to 58\%, and from 47\% to 100\%, respectively, while its bug detection is unaffected.
When we increase \accthreshold beyond 256, increase \allocthreshold beyond 10,000 (\eg 100,000), or decrease the \samplingrate below 10 (\eg 1), \sys behaves similarly to a scenario without any tripwire access control or sampling, which we discuss in \S\ref{sec:discussion}. \sys's bug detection capability drops only when \accthreshold is less than 32 (\eg 93\% for an \accthreshold of 4). 
As we discuss \juliet's limitations in \S\ref{sec:discussion},
\sys's bug detection capability remains $\sim$$97\%$ even with a larger \samplingrate.

In general, while a lower \samplingrate has a significant impact on \sys's run-time overhead, its performance is stable, \ie between 40\% and 60\% as we vary the values
of \accthreshold, \allocthreshold, and \samplingrate.

\takeaway{\sys achieves high bug detection capability with low run-time overhead by using reasonable values for its parameters (\eg \samplingrate $\ge$$100$, \accthreshold $\le$$256$).}

\section{Limitations}
\label{sec:discussion}

\pgheading{\tripwirecap Access Control and Sampling} As we describe in \S\ref{sec:design}, \sys relies on \tripwire access control and sampling to amortize the overhead of additional software checks. Without \tripwire access control and sampling, some applications would suffer from significant run-time overhead, \eg up to $5\times$ and $15\times$ on \bench{perlbench\_r}, respectively.

\revision{\pgheading{Adjusting \sysns’s Parameters for Fuzzing} \sys supports configuring its parameters dynamically to balance between detection strength and performance overhead (\S\ref{sec:design}). Automatic adjustment of this detection-performance tradeoff across many executions of fuzzing would be valuable future work.}

\pgheading{False Negatives}
\sys may suffer from false negatives due to the end of a function and tag collisions. %
In case of a \fault at the end of a function, \ie the \exceptioninsn is followed by a \bench{RET} instruction, \sys could not set the \trap to the \bench{RET} instruction. Consequently, as \sys does not set the \tripwire back to the \shortgranule, it may lead to some potential false negatives.
Also, \sys may increase the possibility of \mte's tag collision. For example, if a pointer's \addrtag 
matches with
\lastbits of a \granule, \sys still permits the memory access even if the access was a true \violation. Treating such \granule as a \shortgranule, \sys still limits the number of addressable bytes to this \granule.

\pgheading{\juliet} Most of the benchmarks in the \juliet are 
micro-benchmarks with only one or two
memory allocations.
Due to such a few memory allocations, \sys does not use sampling from the start of the program.
For example, if \sys enables sampling from the beginning, omitting {\tripwire}s for one out of two {\shortgranule}s, \sys may miss the {\shortgranule} causing the \violation.
Consequently, while sampling from the beginning, \sys fails to detect 10\% of heap-based buffer overflow bugs in \juliet that allocate only a few {\shortgranule}s. 
Updating \juliet to represent realistic memory allocations~\cite{zhou2024characterizing, mteefficient} would be a valuable future work.

\revision{\pgheading{Reducing \sysns’s Overhead Even Further} \sys incurs \specintsys overhead primarily due to \mte’s $\sim12\%$ overhead. Reducing \mte’s overhead even further would require micro-architectural and kernel-level modifications~\cite{noh2026arm, kaushik2025optimized}.}

\pgheading{Side-Channel Attacks} Recent works~\cite{kim2025tiktag, stickytags} demonstrate that the production implementation of ARM \mte (\eg \device) is vulnerable to speculative execution attacks. Consequently, preventing speculative execution attacks on \mte and \mte-enabled systems~\cite{stickytags, mtsan, momeu2025iubik, you2025bastag} using persistent tags~\cite{stickytags} or memory integrity enforcement~\cite{applemie} would be valuable future work.

\section{Related Work}
\label{sec:related}

We classify existing related work between software (\S\ref{sec:related-software}) and hardware (\S\ref{sec:related-hardware}) techniques and compare them against \sys in Table~\ref{tab:comparison-related}.

\begin{table}[t]
\centering
\caption{Comparison of \sys against prior work.} %
\label{tab:comparison-related}
\begin{tabular}{c|c|c|c} 
\toprule
\textbf{Proposal} & \textbf{Granularity} & \textbf{\makecell[c]{Commodity\\ Deployable}} & \textbf{Overhead}\\ 
\midrule
\makecell[c]{Software-Based\\ In-House\\ Techniques} & Byte-Level & Easy & Usually High \\\hline
\makecell[c]{Software-Based\\ In-Production\\ Techniques} & Page-Level & Easy & Low \\\hline
\makecell[c]{Hardware-Based\\In-Production\\ Techniques} & Varies & Relatively Difficult & Usually Low \\\hline
\textbf{\sys} & \textbf{Byte-Level} & \textbf{Easy} & \textbf{Low} \\
\bottomrule
\end{tabular}\\
\end{table}

\subsection{Software-Based Techniques}
\label{sec:related-software}

\pgheading{In-house testing techniques} Software-based memory safety sanitizers are widely-adopted to analyze {\violation}s %
during in-house testing.
Such memory safety sanitizers could be location-based~\cite{asan, hwasan, giantsan, asan--, zhang2021sanrazor, wagner2015asap, nethercote2007valgrind, dinesh2020retrowrite, fioraldi2020qasan} or pointer-based~\cite{nagarakatte2009softbound, nagarakatte2010cets, duck2016lowfat, duck2017stacklowfat, yu2024shadowbound, farkhani2021ptauth, li2022pacmem}.
Among them, ASAN~\cite{asan}, one of the most popular location-based sanitizers, uses redzones and a quarantine list to detect \oob and {\uaf}s, respectively.
Several techniques~\cite{hwasan, asan--, giantsan, whitepaper, zhang2021sanrazor} 
improve upon
ASAN~\cite{overheadasan} %
by avoiding redundant queries~\cite{asan--,zhang2021sanrazor}, segment folding~\cite{giantsan}, or hardware-assisted tagging~\cite{hwasan,whitepaper}. Overall, these software-based techniques are easy to deploy in commodity hardware. They also detect memory safety bugs at byte granularity.
Unfortunately, these software-based techniques typically incur high overhead, slowing down in-house testing, such as fuzzing campaigns. As fuzzing is computationally intensive, %
developers avoid these software-based techniques for many campaigns~\cite{sanddecouple}.
Instead, these campaigns rely on crashes to identify malicious inputs~\cite{ivysyn}, ignoring salient bugs~\cite{fuzzan}. %
\sys addresses this key limitation by enabling low-overhead byte-granular overflow detection.

\pgheading{In-production deployment techniques} Prior work~\cite{serebryany2024gwp, wagner2015asap} also aim to deploy memory safety sanitizers in production via sampling, notably \gwpasan~\cite{serebryany2024gwp}, which Google uses for its various products.
Although \sys uses a similar sampling algorithm to \gwpasan, their design principles are very different. While \gwpasan targets in-production use cases, \sys's primary use case is in-house testing. Therefore, while \gwpasan detects overflows for sampled memory allocations at page granularity, \sys enables byte-granular overflow detection for sampled memory allocations. For unsampled memory allocations, \sys still detects overflows at 16-byte granularity with \mte, while \gwpasan disables any detection.
In general, these sampling-based memory sanitizers target in-production deployment, thus trading their bug detection capability, including granularity, for better performance.

Finally, recent works leverage ARM \mte for sanitization~\cite{mtsan, stickytags, hager2024dmti, chen2024hemate, liljestrand2022color, memtagsan} and  isolation~\cite{momeu2025iubik,kim2024petal,mckee2022hakc,seo2023sfitag,lim2024safebpf, you2025bastag, kha2023capacity}.
\sys improves the effectiveness of these techniques by addressing \mte's coarse precision due to the 16-byte tag granularity.

\subsection{Hardware-Based Techniques}
\label{sec:related-hardware}

Prior work~\cite{watson2015cheri, capstone, yu2025caplification, cornucopia, cctag, devietti2008hardbound, DBLP:conf/osdi/ZeldovichKDK08, timberv, whitepaper,oracleadi, califorms, bogo, intelmpx, safemem, rest} proposes many hardware designs to detect {\violation}s for in-production deployment, including capability-based architectures~\cite{watson2015cheri, capstone, yu2025caplification, cornucopia}, tagged architectures~\cite{cctag, DBLP:conf/osdi/ZeldovichKDK08, timberv, whitepaper, oracleadi, cctag}, hardware bounds checking~\cite{devietti2008hardbound, intelmpx, bogo}, cache-line metadata~\cite{califorms, rest}, and ECC-based techniques~\cite{safemem}. While some of them~\cite{watson2015cheri, capstone, devietti2008hardbound, califorms, intelmpx, bogo} can detect {\violation}s at byte granularity, they remain academic prototypes or have been deprecated~\cite{intelmpx}. For the few~\cite{whitepaper, oracleadi, libmpk} that have commodity implementations, they only have page~\cite{libmpk}, cache-line~\cite{oracleadi} or 16-byte~\cite{whitepaper} granularity.
Altogether, while these hardware-based techniques offer low %
run-time overhead, they either suffer from imprecise granularity or are challenging to deploy with commodity hardware.
In contrast, \sys enables byte-granular detection of {\violation}s in unmodified binaries directly on a commodity implementation of ARM \mte.

Recent hardware proposals~\cite{unterguggenberger2023multitag, mteefficient, song2024parallel, hager2024dmti} improve \mte's tag collision rate~\cite{unterguggenberger2023multitag}, \tagstore overhead~\cite{mteefficient}, performance~\cite{song2024parallel}, and granularity~\cite{hager2024dmti}. %
Specifically, DMTI~\cite{hager2024dmti} improves \mte's tag granularity by %
assigning every memory byte a 4-bit \mte tag in hardware, introducing large \tagstore overhead. %
On the other hand, \sys %
avoids such \tagstore overhead by storing the metadata inside unused padding bytes in the \shortgranule, while still detecting {\violation}s at byte granularity.

\section{Conclusion}
\label{sec:conclusion}

{\violationcap}s remain one of the leading causes of software vulnerabilities. With hardware support, \mte accelerates detecting such bugs, %
but fails to detect \intragranule \oob due to its 16-byte \taggranule.
\revision{We propose \sys to detect {\violation}s probabilistically in unmodified \mte-enabled binaries at byte granularity, addressing \intragranule \oob in real hardware for the first time.}

\section*{Acknowledgments}
\label{sec:acknowledgment}

We thank the anonymous reviewers for their insightful feedback. This work was supported in part by the Center for Ubiquitous Connectivity (CUbiC), sponsored by Semiconductor Research Corporation (SRC) and Defense Advanced Research Projects Agency (DARPA) under the JUMP 2.0 program. This work was partially supported by the Google Cyber NYC Institutional program. We thank the National Science Foundation’s Chameleon Cloud~\cite{chameleon} for providing compute nodes on which we run experiments to obtain some results presented in this paper. We also thank Kostya Serebryany from Tesla, Evgenii Stepanov from Google, and Simha Sethumadhavan from Columbia University for helpful discussions.

\section*{Ethics Considerations}

None.

\bibliographystyle{IEEEtran}
\bibliography{refs}

\begin{thebibliography}{100}
\providecommand{\url}[1]{#1}
\csname url@samestyle\endcsname
\providecommand{\newblock}{\relax}
\providecommand{\bibinfo}[2]{#2}
\providecommand{\BIBentrySTDinterwordspacing}{\spaceskip=0pt\relax}
\providecommand{\BIBentryALTinterwordstretchfactor}{4}
\providecommand{\BIBentryALTinterwordspacing}{\spaceskip=\fontdimen2\font plus
\BIBentryALTinterwordstretchfactor\fontdimen3\font minus \fontdimen4\font\relax}
\providecommand{\BIBforeignlanguage}[2]{{%
\expandafter\ifx\csname l@#1\endcsname\relax
\typeout{** WARNING: IEEEtran.bst: No hyphenation pattern has been}%
\typeout{** loaded for the language `#1'. Using the pattern for}%
\typeout{** the default language instead.}%
\else
\language=\csname l@#1\endcsname
\fi
#2}}
\providecommand{\BIBdecl}{\relax}
\BIBdecl

\bibitem{mattmiller}
\BIBentryALTinterwordspacing
M.~Miller. (2019) Trends, challenge, and shifts in software vulnerability mitigation. [Online]. Available: \url{https://github.com/Microsoft/MSRC-Security-Research/blob/master/presentations/2019_02_BlueHatIL/2019_01%20-%20BlueHatIL%20-%20Trends%2C%20challenge%2C%20and%20shifts%20in%20software%20vulnerability%20mitigation.pdf}
\BIBentrySTDinterwordspacing

\bibitem{aospmemsafety}
``Memory safety | android open source project,'' \url{https://source.android.com/docs/security/test/memory-safety}, [Online; accessed 1-April-2025].

\bibitem{wang2012improving}
\BIBentryALTinterwordspacing
X.~Wang, H.~Chen, Z.~Jia, N.~Zeldovich, and M.~F. Kaashoek, ``Improving integer security for systems with {KINT},'' in \emph{10th {USENIX} Symposium on Operating Systems Design and Implementation, {OSDI} 2012, Hollywood, CA, USA, October 8-10, 2012}, C.~Thekkath and A.~Vahdat, Eds.\hskip 1em plus 0.5em minus 0.4em\relax {USENIX} Association, 2012, pp. 163--177. [Online]. Available: \url{https://www.usenix.org/conference/osdi12/technical-sessions/presentation/wang}
\BIBentrySTDinterwordspacing

\bibitem{li2023hybrid}
\BIBentryALTinterwordspacing
G.~Li, H.~Zhang, J.~Zhou, W.~Shen, Y.~Sui, and Z.~Qian, ``A hybrid alias analysis and its application to global variable protection in the linux kernel,'' in \emph{32nd {USENIX} Security Symposium, {USENIX} Security 2023, Anaheim, CA, USA, August 9-11, 2023}, J.~A. Calandrino and C.~Troncoso, Eds.\hskip 1em plus 0.5em minus 0.4em\relax {USENIX} Association, 2023, pp. 4211--4228. [Online]. Available: \url{https://www.usenix.org/conference/usenixsecurity23/presentation/li-guoren}
\BIBentrySTDinterwordspacing

\bibitem{zhang2025statically}
\BIBentryALTinterwordspacing
H.~Zhang, J.~Kim, C.~Yuan, Z.~Qian, and T.~Kim, ``Statically discover cross-entry use-after-free vulnerabilities in the linux kernel,'' in \emph{32nd Annual Network and Distributed System Security Symposium, {NDSS} 2025, San Diego, California, USA, February 24-28, 2025}.\hskip 1em plus 0.5em minus 0.4em\relax The Internet Society, 2025. [Online]. Available: \url{https://www.ndss-symposium.org/ndss-paper/statically-discover-cross-entry-use-after-free-vulnerabilities-in-the-linux-kernel/}
\BIBentrySTDinterwordspacing

\bibitem{bae2021rudra}
\BIBentryALTinterwordspacing
Y.~Bae, Y.~Kim, A.~Askar, J.~Lim, and T.~Kim, ``Rudra: Finding memory safety bugs in rust at the ecosystem scale,'' in \emph{{SOSP} '21: {ACM} {SIGOPS} 28th Symposium on Operating Systems Principles, Virtual Event / Koblenz, Germany, October 26-29, 2021}, R.~van Renesse and N.~Zeldovich, Eds.\hskip 1em plus 0.5em minus 0.4em\relax {ACM}, 2021, pp. 84--99. [Online]. Available: \url{https://doi.org/10.1145/3477132.3483570}
\BIBentrySTDinterwordspacing

\bibitem{facebookinfer}
\BIBentryALTinterwordspacing
facebook/infer: A static analyzer for java, c, c++, and objective-c. [Online]. Available: \url{https://github.com/facebook/infer}
\BIBentrySTDinterwordspacing

\bibitem{asan}
\BIBentryALTinterwordspacing
K.~Serebryany, D.~Bruening, A.~Potapenko, and D.~Vyukov, ``Addresssanitizer: {A} fast address sanity checker,'' in \emph{Proceedings of the 2012 {USENIX} Annual Technical Conference, {USENIX} {ATC} 2012, Boston, MA, USA, June 13-15, 2012}, G.~Heiser and W.~C. Hsieh, Eds.\hskip 1em plus 0.5em minus 0.4em\relax {USENIX} Association, 2012, pp. 309--318. [Online]. Available: \url{https://www.usenix.org/conference/atc12/technical-sessions/presentation/serebryany}
\BIBentrySTDinterwordspacing

\bibitem{asan--}
\BIBentryALTinterwordspacing
Y.~Zhang, C.~Pang, G.~Portokalidis, N.~Triandopoulos, and J.~Xu, ``Debloating address sanitizer,'' in \emph{31st {USENIX} Security Symposium, {USENIX} Security 2022, Boston, MA, USA, August 10-12, 2022}, K.~R.~B. Butler and K.~Thomas, Eds.\hskip 1em plus 0.5em minus 0.4em\relax {USENIX} Association, 2022, pp. 4345--4363. [Online]. Available: \url{https://www.usenix.org/conference/usenixsecurity22/presentation/zhang-yuchen}
\BIBentrySTDinterwordspacing

\bibitem{duck2018effectivesan}
\BIBentryALTinterwordspacing
G.~J. Duck and R.~H.~C. Yap, ``Effectivesan: type and memory error detection using dynamically typed {C/C++},'' in \emph{Proceedings of the 39th {ACM} {SIGPLAN} Conference on Programming Language Design and Implementation, {PLDI} 2018, Philadelphia, PA, USA, June 18-22, 2018}, J.~S. Foster and D.~Grossman, Eds.\hskip 1em plus 0.5em minus 0.4em\relax {ACM}, 2018, pp. 181--195. [Online]. Available: \url{https://doi.org/10.1145/3192366.3192388}
\BIBentrySTDinterwordspacing

\bibitem{giantsan}
\BIBentryALTinterwordspacing
H.~Ling, H.~Huang, C.~Wang, Y.~Cai, and C.~Zhang, ``{GIANTSAN:} efficient memory sanitization with segment folding,'' in \emph{Proceedings of the 29th {ACM} International Conference on Architectural Support for Programming Languages and Operating Systems, Volume 2, {ASPLOS} 2024, La Jolla, CA, USA, 27 April 2024- 1 May 2024}, R.~Gupta, N.~B. Abu{-}Ghazaleh, M.~Musuvathi, and D.~Tsafrir, Eds.\hskip 1em plus 0.5em minus 0.4em\relax {ACM}, 2024, pp. 433--449. [Online]. Available: \url{https://doi.org/10.1145/3620665.3640391}
\BIBentrySTDinterwordspacing

\bibitem{zhang2021sanrazor}
\BIBentryALTinterwordspacing
J.~Zhang, S.~Wang, M.~Rigger, P.~He, and Z.~Su, ``{SANRAZOR:} reducing redundant sanitizer checks in {C/C++} programs,'' in \emph{15th {USENIX} Symposium on Operating Systems Design and Implementation, {OSDI} 2021, July 14-16, 2021}, A.~D. Brown and J.~R. Lorch, Eds.\hskip 1em plus 0.5em minus 0.4em\relax {USENIX} Association, 2021, pp. 479--494. [Online]. Available: \url{https://www.usenix.org/conference/osdi21/presentation/zhang}
\BIBentrySTDinterwordspacing

\bibitem{yu2024shadowbound}
\BIBentryALTinterwordspacing
Z.~Yu, G.~Yang, and X.~Xing, ``Shadowbound: Efficient heap memory protection through advanced metadata management and customized compiler optimization,'' in \emph{33rd {USENIX} Security Symposium, {USENIX} Security 2024, Philadelphia, PA, USA, August 14-16, 2024}, D.~Balzarotti and W.~Xu, Eds.\hskip 1em plus 0.5em minus 0.4em\relax {USENIX} Association, 2024. [Online]. Available: \url{https://www.usenix.org/conference/usenixsecurity24/presentation/yu-zheng}
\BIBentrySTDinterwordspacing

\bibitem{nagarakatte2009softbound}
\BIBentryALTinterwordspacing
S.~Nagarakatte, J.~Zhao, M.~M.~K. Martin, and S.~Zdancewic, ``Softbound: highly compatible and complete spatial memory safety for c,'' in \emph{Proceedings of the 2009 {ACM} {SIGPLAN} Conference on Programming Language Design and Implementation, {PLDI} 2009, Dublin, Ireland, June 15-21, 2009}, M.~Hind and A.~Diwan, Eds.\hskip 1em plus 0.5em minus 0.4em\relax {ACM}, 2009, pp. 245--258. [Online]. Available: \url{https://doi.org/10.1145/1542476.1542504}
\BIBentrySTDinterwordspacing

\bibitem{hwasan}
``Hardware-assisted addresssanitizer | android open source project,'' \url{https://source.android.com/docs/security/test/hwasan}, [Online; accessed 1-April-2025].

\bibitem{overheadasan}
``Addresssanitizer | android open source project,'' \url{https://source.android.com/docs/security/test/asan}, [Online; accessed 1-April-2025].

\bibitem{dinesh2020retrowrite}
\BIBentryALTinterwordspacing
S.~Dinesh, N.~Burow, D.~Xu, and M.~Payer, ``Retrowrite: Statically instrumenting {COTS} binaries for fuzzing and sanitization,'' in \emph{2020 {IEEE} Symposium on Security and Privacy, {SP} 2020, San Francisco, CA, USA, May 18-21, 2020}.\hskip 1em plus 0.5em minus 0.4em\relax {IEEE}, 2020, pp. 1497--1511. [Online]. Available: \url{https://doi.org/10.1109/SP40000.2020.00009}
\BIBentrySTDinterwordspacing

\bibitem{nethercote2007valgrind}
\BIBentryALTinterwordspacing
N.~Nethercote and J.~Seward, ``Valgrind: a framework for heavyweight dynamic binary instrumentation,'' pp. 89--100, 2007. [Online]. Available: \url{https://doi.org/10.1145/1250734.1250746}
\BIBentrySTDinterwordspacing

\bibitem{fioraldi2020qasan}
\BIBentryALTinterwordspacing
A.~Fioraldi, D.~C. D'Elia, and L.~Querzoni, ``Fuzzing binaries for memory safety errors with qasan,'' in \emph{{IEEE} Secure Development, SecDev 2020, Atlanta, GA, USA, September 28-30, 2020}.\hskip 1em plus 0.5em minus 0.4em\relax {IEEE}, 2020, pp. 23--30. [Online]. Available: \url{https://doi.org/10.1109/SecDev45635.2020.00019}
\BIBentrySTDinterwordspacing

\bibitem{sanddecouple}
\BIBentryALTinterwordspacing
Z.~Kong, S.~Li, H.~Huang, and Z.~Su, ``Sand: Decoupling sanitization from fuzzing for low overhead,'' in \emph{47th {IEEE/ACM} International Conference on Software Engineering, {ICSE} 2025, Ottawa, ON, Canada, April 26 - May 6, 2025}.\hskip 1em plus 0.5em minus 0.4em\relax {IEEE}, 2025, pp. 255--267. [Online]. Available: \url{https://doi.org/10.1109/ICSE55347.2025.00187}
\BIBentrySTDinterwordspacing

\bibitem{fuzzan}
\BIBentryALTinterwordspacing
Y.~Jeon, W.~Han, N.~Burow, and M.~Payer, ``{FuZZan}: Efficient sanitizer metadata design for fuzzing,'' in \emph{2020 USENIX Annual Technical Conference (USENIX ATC 20)}.\hskip 1em plus 0.5em minus 0.4em\relax USENIX Association, Jul. 2020, pp. 249--263. [Online]. Available: \url{https://www.usenix.org/conference/atc20/presentation/jeon}
\BIBentrySTDinterwordspacing

\bibitem{blogarm}
``Enhanced security through memory tagging extension,'' \url{https://community.arm.com/arm-community-blogs/b/architectures-and-processors-blog/posts/enhanced-security-through-mte}, [Online; accessed 1-April-2025].

\bibitem{androidscudo}
``Scudo | android open source project,'' \url{{https://source.android.com/docs/security/test/scudo}}, [Online; accessed 1-April-2025].

\bibitem{projectzero}
``Project zero: First handset with mte on the market,'' \url{https://googleprojectzero.blogspot.com/2023/11/first-handset-with-mte-on-market.html}, [Online; accessed 1-April-2025].

\bibitem{grapheneosmte}
\BIBentryALTinterwordspacing
Memory corruption bug uncovered by mte fixed by google after grapheneos report. [Online]. Available: \url{https://discuss.grapheneos.org/d/12628-memory-corruption-bug-uncovered-by-mte-fixed-by-google-after-grapheneos-report}
\BIBentrySTDinterwordspacing

\bibitem{sync}
\BIBentryALTinterwordspacing
Synchronous mode (sync). [Online]. Available: \url{https://source.android.com/docs/security/test/memory-safety/arm-mte#sync-mode}
\BIBentrySTDinterwordspacing

\bibitem{juliet}
``Juliet c/c++ 1.3 - nist software assurance reference dataset,'' \url{https://samate.nist.gov/SARD/test-suites/112}, [Online; accessed 1-April-2025].

\bibitem{gnuallocator}
``3.2.2 the gnu allocator,'' \url{https://www.gnu.org/software/libc/manual/html_node/The-GNU-Allocator.html}, [Online; accessed 1-April-2025].

\bibitem{stickytags}
\BIBentryALTinterwordspacing
F.~Gorter, T.~Kroes, H.~Bos, and C.~Giuffrida, ``Sticky tags: Efficient and deterministic spatial memory error mitigation using persistent memory tags,'' in \emph{{IEEE} Symposium on Security and Privacy, {SP} 2024, San Francisco, CA, USA, May 19-23, 2024}.\hskip 1em plus 0.5em minus 0.4em\relax {IEEE}, 2024, pp. 4239--4257. [Online]. Available: \url{https://doi.org/10.1109/SP54263.2024.00263}
\BIBentrySTDinterwordspacing

\bibitem{khan2019huron}
T.~A. Khan, Y.~Zhao, G.~Pokam, B.~Mozafari, and B.~Kasikci, ``Huron: hybrid false sharing detection and repair,'' in \emph{Proceedings of the 40th ACM SIGPLAN Conference on Programming Language Design and Implementation}, 2019, pp. 453--468.

\bibitem{specbenchmark}
\BIBentryALTinterwordspacing
Spec cpu® 2017 benchmark. [Online]. Available: \url{https://www.spec.org/cpu2017/}
\BIBentrySTDinterwordspacing

\bibitem{cveexample}
``Nvd - cve-2024-12084,'' \url{https://nvd.nist.gov/vuln/detail/CVE-2024-12084}, [Online; accessed 1-April-2025].

\bibitem{securitygoogle}
\BIBentryALTinterwordspacing
Rsync: Heap buffer overflow, info leak, server leaks, path traversal and safe links bypass. [Online]. Available: \url{https://github.com/google/security-research/security/advisories/GHSA-p5pg-x43v-mvqj}
\BIBentrySTDinterwordspacing

\bibitem{redhatrating}
``Cve-2024-12084 - red hat customer portal,'' \url{https://access.redhat.com/security/cve/CVE-2024-12084}, [Online; accessed 1-April-2025].

\bibitem{hastings1992purify}
R.~Hastings and B.~Joyce, ``Purify: A tool for detecting memory leaks and access errors in c and c++ programs,'' in \emph{Proceedings of the Winter 1992 USENIX Conference}, pp. 125--138.

\bibitem{efence}
\BIBentryALTinterwordspacing
B.~Perens. Electric fence. [Online]. Available: \url{https://elinux.org/Electric\_Fence}
\BIBentrySTDinterwordspacing

\bibitem{breakpoint}
\BIBentryALTinterwordspacing
Overview: Breakpoints and watchpoints. [Online]. Available: \url{https://developer.arm.com/documentation/dui0446/z/controlling-target-execution/overview--breakpoints-and-watchpoints}
\BIBentrySTDinterwordspacing

\bibitem{geekbench}
``Geekbench 6 - cross-platform benchmark,'' \url{https://www.geekbench.com/}, [Online; accessed 1-April-2025].

\bibitem{memcached}
\BIBentryALTinterwordspacing
memcached - a distributed memory object caching system. [Online]. Available: \url{https://memcached.org/}
\BIBentrySTDinterwordspacing

\bibitem{leveldb}
\BIBentryALTinterwordspacing
google/leveldb. [Online]. Available: \url{https://github.com/google/leveldb}
\BIBentrySTDinterwordspacing

\bibitem{rocksdb}
\BIBentryALTinterwordspacing
facebook/rocksdb. [Online]. Available: \url{https://github.com/facebook/rocksdb}
\BIBentrySTDinterwordspacing

\bibitem{whitepaper}
``Armv8.5-a memory tagging extension white paper,'' \url{https://developer.arm.com/documentation/102925/latest/}, [Online; accessed 1-April-2025].

\bibitem{mteefficient}
\BIBentryALTinterwordspacing
A.~Partap and D.~Boneh, ``Memory tagging: {A} memory efficient design,'' \emph{CoRR}, vol. abs/2209.00307, 2022. [Online]. Available: \url{https://doi.org/10.48550/arXiv.2209.00307}
\BIBentrySTDinterwordspacing

\bibitem{tbi}
``Tagged pointers | android open source project,'' \url{https://source.android.com/docs/security/test/tagged-pointers}, [Online; accessed 1-April-2025].

\bibitem{tagstore}
``Where is the mte tag stored and checked?'' \url{{https://developer.arm.com/documentation/ka005620/1-0/?lang=en}}, [Online; accessed 1-April-2025].

\bibitem{mtemodes}
``Mte modes,'' \url{https://developer.arm.com/documentation/108035/0100/How-does-MTE-work-/MTE-modes}, [Online; accessed 1-April-2025].

\bibitem{loadstoreinsns}
``Arm a-profile a64 instruction set architecture | loads and stores,'' \url{https://developer.arm.com/documentation/ddi0602/2024-09/Index-by-Encoding/Loads-and-Stores?lang=en}, [Online; accessed 1-April-2025].

\bibitem{mtsan}
\BIBentryALTinterwordspacing
X.~Chen, Y.~Shi, Z.~Jiang, Y.~Li, R.~Wang, H.~Duan, H.~Wang, and C.~Zhang, ``Mtsan: {A} feasible and practical memory sanitizer for fuzzing {COTS} binaries,'' in \emph{32nd {USENIX} Security Symposium, {USENIX} Security 2023, Anaheim, CA, USA, August 9-11, 2023}, J.~A. Calandrino and C.~Troncoso, Eds.\hskip 1em plus 0.5em minus 0.4em\relax {USENIX} Association, 2023, pp. 841--858. [Online]. Available: \url{https://www.usenix.org/conference/usenixsecurity23/presentation/chen-xingman}
\BIBentrySTDinterwordspacing

\bibitem{scudodccommit}
P.~Collingbourne, ``scudo: Use dc gzva instruction in storetags(),'' \url{https://github.com/llvm/llvm-project/commit/46c59d91dc7a39cc98be7a68d6dc60f3e8a35df0}, [Online; accessed 1-April-2025].

\bibitem{kernelmte}
``Memory tagging extension (mte) in aarch64 linux,'' \url{http://docs.kernel.org/arch/arm64/memory-tagging-extension.html\#memory-tagging-extension-mte-in-aarch64-linux}, [Online; accessed 1-April-2025].

\bibitem{bootloader}
``Mte bootloader support,'' \url{{https://source.android.com/docs/security/test/memory-safety/bootloader-support}}, [Online; accessed 1-April-2025].

\bibitem{li2022pacmem}
\BIBentryALTinterwordspacing
Y.~Li, W.~Tan, Z.~Lv, S.~Yang, M.~Payer, Y.~Liu, and C.~Zhang, ``Pacmem: Enforcing spatial and temporal memory safety via {ARM} pointer authentication,'' in \emph{Proceedings of the 2022 {ACM} {SIGSAC} Conference on Computer and Communications Security, {CCS} 2022, Los Angeles, CA, USA, November 7-11, 2022}, H.~Yin, A.~Stavrou, C.~Cremers, and E.~Shi, Eds.\hskip 1em plus 0.5em minus 0.4em\relax {ACM}, 2022, pp. 1901--1915. [Online]. Available: \url{https://doi.org/10.1145/3548606.3560598}
\BIBentrySTDinterwordspacing

\bibitem{cwe2024}
``Cwe - 2024 cwe top 25 most dangerous software weaknesses,'' \url{https://cwe.mitre.org/top25/archive/2024/2024_cwe_top25.html}, [Online; accessed 1-April-2025].

\bibitem{serebryany2024gwp}
\BIBentryALTinterwordspacing
K.~Serebryany, C.~Kennelly, M.~Phillips, M.~Denton, M.~Elver, A.~Potapenko, M.~Morehouse, V.~Tsyrklevich, C.~Holler, J.~Lettner, D.~Kilzer, and L.~Brandt, ``Gwp-asan: Sampling-based detection of memory-safety bugs in production,'' in \emph{Proceedings of the 46th International Conference on Software Engineering: Software Engineering in Practice, {ICSE-SEIP} 2024, Lisbon, Portugal, April 14-20, 2024}.\hskip 1em plus 0.5em minus 0.4em\relax {ACM}, 2024, pp. 168--177. [Online]. Available: \url{https://doi.org/10.1145/3639477.3640328}
\BIBentrySTDinterwordspacing

\bibitem{lsan}
``Leaksanitizer - clang 21.0.0git documentation,'' \url{https://clang.llvm.org/docs/LeakSanitizer.html}, [Online; accessed 1-April-2025].

\bibitem{kostyamemtag}
\BIBentryALTinterwordspacing
K.~Serebryany, E.~Stepanov, A.~Shlyapnikov, V.~Tsyrklevich, and D.~Vyukov, ``Memory tagging and how it improves {C/C++} memory safety,'' \emph{CoRR}, vol. abs/1802.09517, 2018. [Online]. Available: \url{http://arxiv.org/abs/1802.09517}
\BIBentrySTDinterwordspacing

\bibitem{mddefinesh}
``rsync/lib/md-defines.h,'' \url{https://github.com/RsyncProject/rsync/blob/9615a2492bbf96bc145e738ebff55bbb91e0bbee/lib/md-defines.h\#L11-L21}, [Online; accessed 1-April-2025].

\bibitem{codersynch}
``rsync/rsync.c,'' \url{https://github.com/RsyncProject/rsync/blob/9615a2492bbf96bc145e738ebff55bbb91e0bbee/rsync.h\#L955-L962}, [Online; accessed 1-April-2025].

\bibitem{senderc}
``rsync/sender.c,'' \url{https://github.com/RsyncProject/rsync/blob/9615a2492bbf96bc145e738ebff55bbb91e0bbee/sender.c\#L96-L100}, [Online; accessed 1-April-2025].

\bibitem{rsync}
``rsync,'' \url{https://rsync.samba.org/}, [Online; accessed 1-April-2025].

\bibitem{kim2025tiktag}
\BIBentryALTinterwordspacing
J.~Kim, J.~Park, S.~Roh, J.~Chung, Y.~Lee, T.~Kim, and B.~Lee, ``Tiktag: Breaking arm's memory tagging extension with speculative execution,'' in \emph{{IEEE} Symposium on Security and Privacy, {SP} 2025, San Francisco, CA, USA, May 12-15, 2025}, M.~Blanton, W.~Enck, and C.~Nita{-}Rotaru, Eds.\hskip 1em plus 0.5em minus 0.4em\relax {IEEE}, 2025, pp. 4063--4081. [Online]. Available: \url{https://doi.org/10.1109/SP61157.2025.00039}
\BIBentrySTDinterwordspacing

\bibitem{kocher2020spectre}
P.~Kocher, J.~Horn, A.~Fogh, D.~Genkin, D.~Gruss, W.~Haas, M.~Hamburg, M.~Lipp, S.~Mangard, T.~Prescher \emph{et~al.}, ``Spectre attacks: Exploiting speculative execution,'' \emph{Communications of the ACM}, vol.~63, no.~7, pp. 93--101, 2020.

\bibitem{lipp2020meltdown}
M.~Lipp, M.~Schwarz, D.~Gruss, T.~Prescher, W.~Haas, J.~Horn, S.~Mangard, P.~Kocher, D.~Genkin, Y.~Yarom \emph{et~al.}, ``Meltdown: Reading kernel memory from user space,'' \emph{Communications of the ACM}, vol.~63, no.~6, pp. 46--56, 2020.

\bibitem{debugregs}
\BIBentryALTinterwordspacing
About the debug register interface. [Online]. Available: \url{https://developer.arm.com/documentation/ddi0379/a/Debug-Register-Interfaces/About-the-Debug-Register-Interface?lang=en}
\BIBentrySTDinterwordspacing

\bibitem{signalhandler}
``Signal handling (the gnu c library),'' \url{https://www.gnu.org/software/libc/manual/html_node/Signal-Handling.html}, [Online; accessed 1-April-2025].

\bibitem{envvar}
\BIBentryALTinterwordspacing
Environmentvariables - community help wiki. [Online]. Available: \url{https://help.ubuntu.com/community/EnvironmentVariables}
\BIBentrySTDinterwordspacing

\bibitem{glibc}
``The gnu c library - gnu project - free software foundation,'' \url{https://www.gnu.org/software/libc/}, [Online; accessed 1-April-2025].

\bibitem{strcpy}
``glibc/sysdeps/aarch64/strcpy.s,'' \url{https://github.com/bminor/glibc/blob/ce2f26a22e6b6f5c108d156afd9b43a452bb024c/sysdeps/aarch64/strcpy.S}, [Online; accessed 1-April-2025].

\bibitem{compatiblestrcpy}
``aarch64: Mte compatible strcpy,'' \url{https://github.com/bminor/glibc/commit/bb2c12aecbd26a8d29f63b51b80b7c84e65d1818}, [Online; accessed 1-April-2025].

\bibitem{ucontext}
\BIBentryALTinterwordspacing
System v contexts (the gnu c library). [Online]. Available: \url{https://www.gnu.org/software/libc/manual/html_node/System-V-contexts.html}
\BIBentrySTDinterwordspacing

\bibitem{momeu2025iubik}
\BIBentryALTinterwordspacing
M.~Momeu, A.~J. Gaidis, J.~v.~d. Heidt, and V.~P. Kemerlis, ``{IUBIK:} isolating user bytes in commodity operating system kernels via memory tagging extensions,'' in \emph{{IEEE} Symposium on Security and Privacy, {SP} 2025, San Francisco, CA, USA, May 12-15, 2025}, M.~Blanton, W.~Enck, and C.~Nita{-}Rotaru, Eds.\hskip 1em plus 0.5em minus 0.4em\relax {IEEE}, 2025, pp. 867--885. [Online]. Available: \url{https://doi.org/10.1109/SP61157.2025.00135}
\BIBentrySTDinterwordspacing

\bibitem{termux}
``Termux | the main termux site and help pages.'' \url{https://termux.dev/en/}, [Online; accessed 1-April-2025].

\bibitem{mantermux}
``Getting started - termux wiki,'' \url{https://wiki.termux.com/wiki/Getting_started}, [Online; accessed 1-April-2025].

\bibitem{chroot}
``chroot(2) - linux manual page,'' \url{https://man7.org/linux/man-pages/man2/chroot.2.html}, [Online; accessed 1-April-2025].

\bibitem{clang14}
\BIBentryALTinterwordspacing
Clang 14.0.0 documentation. [Online]. Available: \url{https://releases.llvm.org/14.0.0/tools/clang/docs/index.html}
\BIBentrySTDinterwordspacing

\bibitem{hazimeh2020magma}
\BIBentryALTinterwordspacing
A.~Hazimeh, A.~Herrera, and M.~Payer, ``Magma: {A} ground-truth fuzzing benchmark,'' \emph{Proc. {ACM} Meas. Anal. Comput. Syst.}, vol.~4, no.~3, pp. 49:1--49:29, 2020. [Online]. Available: \url{https://doi.org/10.1145/3428334}
\BIBentrySTDinterwordspacing

\bibitem{unterguggenberger2023multitag}
\BIBentryALTinterwordspacing
M.~Unterguggenberger, D.~Schrammel, P.~Nasahl, R.~Schilling, L.~Lamster, and S.~Mangard, ``Multi-tag: {A} hardware-software co-design for memory safety based on multi-granular memory tagging,'' in \emph{Proceedings of the 2023 {ACM} Asia Conference on Computer and Communications Security, {ASIA} {CCS} 2023, Melbourne, VIC, Australia, July 10-14, 2023}, J.~K. Liu, Y.~Xiang, S.~Nepal, and G.~Tsudik, Eds.\hskip 1em plus 0.5em minus 0.4em\relax {ACM}, 2023, pp. 177--189. [Online]. Available: \url{https://doi.org/10.1145/3579856.3590331}
\BIBentrySTDinterwordspacing

\bibitem{thunderx2}
\BIBentryALTinterwordspacing
Thunderx2 - cavium - wikichip. [Online]. Available: \url{https://en.wikichip.org/wiki/cavium/thunderx2}
\BIBentrySTDinterwordspacing

\bibitem{chameleon}
\BIBentryALTinterwordspacing
K.~Keahey, J.~Anderson, Z.~Zhen, P.~Riteau, P.~Ruth, D.~Stanzione, M.~Cevik, J.~Colleran, H.~S. Gunawi, C.~Hammock, J.~Mambretti, A.~Barnes, F.~Halbah, A.~Rocha, and J.~Stubbs, ``Lessons learned from the chameleon testbed,'' in \emph{2020 USENIX Annual Technical Conference (USENIX ATC 20)}.\hskip 1em plus 0.5em minus 0.4em\relax USENIX Association, Jul. 2020, pp. 219--233. [Online]. Available: \url{https://www.usenix.org/conference/atc20/presentation/keahey}
\BIBentrySTDinterwordspacing

\bibitem{avoidruncpu}
\BIBentryALTinterwordspacing
Avoiding runcpu - cpu 2017. [Online]. Available: \url{https://www.spec.org/cpu2017/Docs/runcpu-avoidance.html}
\BIBentrySTDinterwordspacing

\bibitem{previewgeekbench}
``Preview versions - geekbench,'' \url{https://www.geekbench.com/preview/}, [Online; accessed 1-April-2025].

\bibitem{playstoregeekbench}
``Geekbench 6 - apps on google play,'' \url{https://play.google.com/store/apps/details?id=com.primatelabs.geekbench6&hl=en_US}, [Online; accessed 1-April-2025].

\bibitem{appstoregeekbench}
``Geekbench 6 on the app store,'' \url{https://apps.apple.com/us/app/geekbench-6/id1565728895}, [Online; accessed 1-April-2025].

\bibitem{seward2005using}
\BIBentryALTinterwordspacing
J.~Seward and N.~Nethercote, ``Using valgrind to detect undefined value errors with bit-precision,'' in \emph{Proceedings of the 2005 {USENIX} Annual Technical Conference, April 10-15, 2005, Anaheim, CA, {USA}}.\hskip 1em plus 0.5em minus 0.4em\relax {USENIX}, 2005, pp. 17--30. [Online]. Available: \url{http://www.usenix.org/events/usenix05/tech/general/seward.html}
\BIBentrySTDinterwordspacing

\bibitem{mc-benchmark}
\BIBentryALTinterwordspacing
antirez/mc-benchmark. [Online]. Available: \url{https://github.com/antirez/mc-benchmark}
\BIBentrySTDinterwordspacing

\bibitem{ldb-dbbench}
\BIBentryALTinterwordspacing
leveldb/benchmarks/db\_bench.cc. [Online]. Available: \url{https://github.com/google/leveldb/blob/main/benchmarks/db\_bench.cc}
\BIBentrySTDinterwordspacing

\bibitem{rdb-dbbench}
\BIBentryALTinterwordspacing
rocksdb/tools/db\_bench\_tool.cc. [Online]. Available: \url{https://github.com/facebook/rocksdb/blob/main/tools/db\_bench\_tool.cc}
\BIBentrySTDinterwordspacing

\bibitem{ugur-one-profile-fits-all-osr-2022}
M.~Ugur, C.~Jiang, A.~Erf, T.~A. Khan, and B.~Kasikci, ``One profile fits all: Profile-guided linux kernel optimizations for data center applications,'' \emph{ACM SIGOPS Operating Systems Review}, vol.~56, no.~1, pp. 26--33, Jun. 2022.

\bibitem{magmagithub}
\BIBentryALTinterwordspacing
Release v1.2.1, hexhive/magma. [Online]. Available: \url{https://github.com/HexHive/magma/tree/v1.2.1}
\BIBentrySTDinterwordspacing

\bibitem{fioraldi2020afl++}
\BIBentryALTinterwordspacing
A.~Fioraldi, D.~C. Maier, H.~Ei{\ss}feldt, and M.~Heuse, ``{AFL++} : Combining incremental steps of fuzzing research,'' in \emph{14th {USENIX} Workshop on Offensive Technologies, {WOOT} 2020, August 11, 2020}, Y.~Yarom and S.~Zennou, Eds.\hskip 1em plus 0.5em minus 0.4em\relax {USENIX} Association, 2020. [Online]. Available: \url{https://www.usenix.org/conference/woot20/presentation/fioraldi}
\BIBentrySTDinterwordspacing

\bibitem{persistentmode}
\BIBentryALTinterwordspacing
llvm\_mode persistent mode. [Online]. Available: \url{https://github.com/AFLplusplus/AFLplusplus/blob/stable/instrumentation/README.persistent_mode.md}
\BIBentrySTDinterwordspacing

\bibitem{ampereone}
\BIBentryALTinterwordspacing
Ampereone product brief. [Online]. Available: \url{https://amperecomputing.com/briefs/ampereone-family-product-brief}
\BIBentrySTDinterwordspacing

\bibitem{kaushik2025optimized}
S.~Kaushik, M.~Madhav, N.~Aboulenein, J.~Bessette, S.~Brahmadathan, B.~Chaffin, M.~Erler, S.~Jourdan, T.~Maciukenas, R.~Masti \emph{et~al.}, ``Optimized memory tagging on ampereone processors,'' \emph{arXiv preprint arXiv:2511.17773}, 2025.

\bibitem{xu2017designing}
\BIBentryALTinterwordspacing
W.~Xu, S.~Kashyap, C.~Min, and T.~Kim, ``Designing new operating primitives to improve fuzzing performance,'' in \emph{Proceedings of the 2017 {ACM} {SIGSAC} Conference on Computer and Communications Security, {CCS} 2017, Dallas, TX, USA, October 30 - November 03, 2017}, B.~Thuraisingham, D.~Evans, T.~Malkin, and D.~Xu, Eds.\hskip 1em plus 0.5em minus 0.4em\relax {ACM}, 2017, pp. 2313--2328. [Online]. Available: \url{https://doi.org/10.1145/3133956.3134046}
\BIBentrySTDinterwordspacing

\bibitem{zhou2024characterizing}
\BIBentryALTinterwordspacing
Z.~Zhou, V.~Gogte, N.~Vaish, C.~Kennelly, P.~Xia, S.~Kanev, T.~Moseley, C.~Delimitrou, and P.~Ranganathan, ``Characterizing a memory allocator at warehouse scale,'' in \emph{Proceedings of the 29th {ACM} International Conference on Architectural Support for Programming Languages and Operating Systems, Volume 3, {ASPLOS} 2024, La Jolla, CA, USA, 27 April 2024- 1 May 2024}, R.~Gupta, N.~B. Abu{-}Ghazaleh, M.~Musuvathi, and D.~Tsafrir, Eds.\hskip 1em plus 0.5em minus 0.4em\relax {ACM}, 2024, pp. 192--206. [Online]. Available: \url{https://doi.org/10.1145/3620666.3651350}
\BIBentrySTDinterwordspacing

\bibitem{noh2026arm}
T.~Noh, Y.~Wang, T.~Garfinkel, M.~Madhav, D.~Moghimi, M.~Erez, and S.~Narayan, ``Arm mte performance in practice,'' in \emph{Usenix Security Symposium}, 2026.

\bibitem{you2025bastag}
\BIBentryALTinterwordspacing
J.~You, J.~Seo, K.~Lee, Y.~Cho, and Y.~Paek, ``Bastag: Byte-level access control on shared memory using arm memory tagging extension,'' in \emph{Proceedings of the 32nd ACM Conference on Computer and Communications Security (CCS '25)}.\hskip 1em plus 0.5em minus 0.4em\relax New York, NY, USA: Association for Computing Machinery, 2025. [Online]. Available: \url{https://junseungyou.github.io/assets/bastag.pdf}
\BIBentrySTDinterwordspacing

\bibitem{applemie}
\BIBentryALTinterwordspacing
Memory integrity enforcement: A complete vision for memory safety in apple devices. [Online]. Available: \url{https://security.apple.com/blog/memory-integrity-enforcement/}
\BIBentrySTDinterwordspacing

\bibitem{wagner2015asap}
\BIBentryALTinterwordspacing
J.~Wagner, V.~Kuznetsov, G.~Candea, and J.~Kinder, ``High system-code security with low overhead,'' in \emph{2015 {IEEE} Symposium on Security and Privacy, {SP} 2015, San Jose, CA, USA, May 17-21, 2015}.\hskip 1em plus 0.5em minus 0.4em\relax {IEEE} Computer Society, 2015, pp. 866--879. [Online]. Available: \url{https://doi.org/10.1109/SP.2015.58}
\BIBentrySTDinterwordspacing

\bibitem{nagarakatte2010cets}
\BIBentryALTinterwordspacing
S.~Nagarakatte, J.~Zhao, M.~M.~K. Martin, and S.~Zdancewic, ``{CETS:} compiler enforced temporal safety for {C},'' in \emph{Proceedings of the 9th International Symposium on Memory Management, {ISMM} 2010, Toronto, Ontario, Canada, June 5-6, 2010}, J.~Vitek and D.~Lea, Eds.\hskip 1em plus 0.5em minus 0.4em\relax {ACM}, 2010, pp. 31--40. [Online]. Available: \url{https://doi.org/10.1145/1806651.1806657}
\BIBentrySTDinterwordspacing

\bibitem{duck2016lowfat}
\BIBentryALTinterwordspacing
G.~J. Duck and R.~H.~C. Yap, ``Heap bounds protection with low fat pointers,'' in \emph{Proceedings of the 25th International Conference on Compiler Construction, {CC} 2016, Barcelona, Spain, March 12-18, 2016}, A.~Zaks and M.~V. Hermenegildo, Eds.\hskip 1em plus 0.5em minus 0.4em\relax {ACM}, 2016, pp. 132--142. [Online]. Available: \url{https://doi.org/10.1145/2892208.2892212}
\BIBentrySTDinterwordspacing

\bibitem{duck2017stacklowfat}
\BIBentryALTinterwordspacing
G.~J. Duck, R.~H.~C. Yap, and L.~Cavallaro, ``Stack bounds protection with low fat pointers,'' in \emph{24th Annual Network and Distributed System Security Symposium, {NDSS} 2017, San Diego, California, USA, February 26 - March 1, 2017}.\hskip 1em plus 0.5em minus 0.4em\relax The Internet Society, 2017. [Online]. Available: \url{https://www.ndss-symposium.org/ndss2017/ndss-2017-programme/stack-object-protection-low-fat-pointers/}
\BIBentrySTDinterwordspacing

\bibitem{farkhani2021ptauth}
\BIBentryALTinterwordspacing
R.~M. Farkhani, M.~Ahmadi, and L.~Lu, ``Ptauth: Temporal memory safety via robust points-to authentication,'' in \emph{30th {USENIX} Security Symposium, {USENIX} Security 2021, August 11-13, 2021}, M.~D. Bailey and R.~Greenstadt, Eds.\hskip 1em plus 0.5em minus 0.4em\relax {USENIX} Association, 2021, pp. 1037--1054. [Online]. Available: \url{https://www.usenix.org/conference/usenixsecurity21/presentation/mirzazade}
\BIBentrySTDinterwordspacing

\bibitem{ivysyn}
\BIBentryALTinterwordspacing
N.~Christou, D.~Jin, V.~Atlidakis, B.~Ray, and V.~P. Kemerlis, ``Ivysyn: Automated vulnerability discovery in deep learning frameworks,'' in \emph{32nd {USENIX} Security Symposium, {USENIX} Security 2023, Anaheim, CA, USA, August 9-11, 2023}, J.~A. Calandrino and C.~Troncoso, Eds.\hskip 1em plus 0.5em minus 0.4em\relax {USENIX} Association, 2023, pp. 2383--2400. [Online]. Available: \url{https://www.usenix.org/conference/usenixsecurity23/presentation/christou}
\BIBentrySTDinterwordspacing

\bibitem{hager2024dmti}
\BIBentryALTinterwordspacing
A.~Hager{-}Clukas and K.~Hohentanner, ``{DMTI:} accelerating memory error detection in precompiled {C/C++} binaries with {ARM} memory tagging extension,'' in \emph{Proceedings of the 19th {ACM} Asia Conference on Computer and Communications Security, {ASIA} {CCS} 2024, Singapore, July 1-5, 2024}, J.~Zhou, T.~Q.~S. Quek, D.~Gao, and A.~A. C{\'{a}}rdenas, Eds.\hskip 1em plus 0.5em minus 0.4em\relax {ACM}, 2024. [Online]. Available: \url{https://doi.org/10.1145/3634737.3637655}
\BIBentrySTDinterwordspacing

\bibitem{chen2024hemate}
\BIBentryALTinterwordspacing
Y.~Chen and S.~Li, ``Hemate: Enhancing heap security through isolating primitive types with arm memory tagging extension,'' in \emph{Proceedings of the 19th International Conference on Availability, Reliability and Security, {ARES} 2024, Vienna, Austria, 30 July 2024 - 2 August 2024}.\hskip 1em plus 0.5em minus 0.4em\relax {ACM}, 2024, pp. 30:1--30:11. [Online]. Available: \url{https://doi.org/10.1145/3664476.3664492}
\BIBentrySTDinterwordspacing

\bibitem{liljestrand2022color}
\BIBentryALTinterwordspacing
H.~Liljestrand, C.~C. Perez, R.~Denis{-}Courmont, J.~Ekberg, and N.~Asokan, ``Color my world: Deterministic tagging for memory safety,'' \emph{CoRR}, vol. abs/2204.03781, 2022. [Online]. Available: \url{https://doi.org/10.48550/arXiv.2204.03781}
\BIBentrySTDinterwordspacing

\bibitem{memtagsan}
``Memtagsanitizer - llvm 21.0.0git documentation,'' \url{https://llvm.org/docs/MemTagSanitizer.html}, [Online; accessed 1-April-2025].

\bibitem{kim2024petal}
\BIBentryALTinterwordspacing
J.~Kim, J.~Park, Y.~Lee, C.~Song, T.~Kim, and B.~Lee, ``Petal: Ensuring access control integrity against data-only attacks on linux,'' in \emph{Proceedings of the 2024 on {ACM} {SIGSAC} Conference on Computer and Communications Security, {CCS} 2024, Salt Lake City, UT, USA, October 14-18, 2024}, B.~Luo, X.~Liao, J.~Xu, E.~Kirda, and D.~Lie, Eds.\hskip 1em plus 0.5em minus 0.4em\relax {ACM}, 2024, pp. 2919--2933. [Online]. Available: \url{https://doi.org/10.1145/3658644.3690184}
\BIBentrySTDinterwordspacing

\bibitem{mckee2022hakc}
\BIBentryALTinterwordspacing
D.~P. McKee, Y.~Giannaris, C.~Ortega, H.~E. Shrobe, M.~Payer, H.~Okhravi, and N.~Burow, ``Preventing kernel hacks with hakcs,'' in \emph{29th Annual Network and Distributed System Security Symposium, {NDSS} 2022, San Diego, California, USA, April 24-28, 2022}.\hskip 1em plus 0.5em minus 0.4em\relax The Internet Society, 2022. [Online]. Available: \url{https://www.ndss-symposium.org/ndss-paper/auto-draft-257/}
\BIBentrySTDinterwordspacing

\bibitem{seo2023sfitag}
\BIBentryALTinterwordspacing
J.~Seo, J.~You, Y.~Cho, Y.~Cho, D.~Kwon, and Y.~Paek, ``Sfitag: Efficient software fault isolation with memory tagging for {ARM} kernel extensions,'' in \emph{Proceedings of the 2023 {ACM} Asia Conference on Computer and Communications Security, {ASIA} {CCS} 2023, Melbourne, VIC, Australia, July 10-14, 2023}, J.~K. Liu, Y.~Xiang, S.~Nepal, and G.~Tsudik, Eds.\hskip 1em plus 0.5em minus 0.4em\relax {ACM}, 2023, pp. 469--480. [Online]. Available: \url{https://doi.org/10.1145/3579856.3590341}
\BIBentrySTDinterwordspacing

\bibitem{lim2024safebpf}
\BIBentryALTinterwordspacing
S.~Lim, T.~Prasad, X.~Han, and T.~Pasquier, ``Safebpf: Hardware-assisted defense-in-depth for ebpf kernel extensions,'' in \emph{Proceedings of the 2024 on Cloud Computing Security Workshop, {CCSW} 2024, Salt Lake City, UT, USA, October 14-18, 2024}, A.~P. Fournaris and P.~Palmieri, Eds.\hskip 1em plus 0.5em minus 0.4em\relax {ACM}, 2024, pp. 80--94. [Online]. Available: \url{https://doi.org/10.1145/3689938.3694781}
\BIBentrySTDinterwordspacing

\bibitem{kha2023capacity}
\BIBentryALTinterwordspacing
K.~D. Duy, K.~Cho, T.~Noh, and H.~Lee, ``Capacity: Cryptographically-enforced in-process capabilities for modern {ARM} architectures,'' in \emph{Proceedings of the 2023 {ACM} {SIGSAC} Conference on Computer and Communications Security, {CCS} 2023, Copenhagen, Denmark, November 26-30, 2023}, W.~Meng, C.~D. Jensen, C.~Cremers, and E.~Kirda, Eds.\hskip 1em plus 0.5em minus 0.4em\relax {ACM}, 2023, pp. 874--888. [Online]. Available: \url{https://doi.org/10.1145/3576915.3623079}
\BIBentrySTDinterwordspacing

\bibitem{watson2015cheri}
\BIBentryALTinterwordspacing
R.~N.~M. Watson, J.~Woodruff, P.~G. Neumann, S.~W. Moore, J.~Anderson, D.~Chisnall, N.~H. Dave, B.~Davis, K.~Gudka, B.~Laurie, S.~J. Murdoch, R.~M. Norton, M.~Roe, S.~D. Son, and M.~Vadera, ``{CHERI:} {A} hybrid capability-system architecture for scalable software compartmentalization,'' in \emph{2015 {IEEE} Symposium on Security and Privacy, {SP} 2015, San Jose, CA, USA, May 17-21, 2015}.\hskip 1em plus 0.5em minus 0.4em\relax {IEEE} Computer Society, 2015, pp. 20--37. [Online]. Available: \url{https://doi.org/10.1109/SP.2015.9}
\BIBentrySTDinterwordspacing

\bibitem{capstone}
\BIBentryALTinterwordspacing
J.~Z. Yu, C.~Watt, A.~Badole, T.~E. Carlson, and P.~Saxena, ``Capstone: {A} capability-based foundation for trustless secure memory access,'' in \emph{32nd {USENIX} Security Symposium, {USENIX} Security 2023, Anaheim, CA, USA, August 9-11, 2023}, J.~A. Calandrino and C.~Troncoso, Eds.\hskip 1em plus 0.5em minus 0.4em\relax {USENIX} Association, 2023, pp. 787--804. [Online]. Available: \url{https://www.usenix.org/conference/usenixsecurity23/presentation/yu-jason}
\BIBentrySTDinterwordspacing

\bibitem{yu2025caplification}
J.~Z. Yu, M.~Li, A.~Badole, T.~E. Carlson, M.~Swift, and P.~Saxena, ``Caplification: Bridging capability-aware and capability-oblivious software,'' in \emph{Proceedings of the 30th ACM Symposium on Access Control Models and Technologies}, 2025, pp. 33--44.

\bibitem{cornucopia}
\BIBentryALTinterwordspacing
N.~W. Filardo, B.~F. Gutstein, J.~Woodruff, S.~Ainsworth, L.~Paul{-}Trifu, B.~Davis, H.~Xia, E.~T. Napierala, A.~Richardson, J.~Baldwin, D.~Chisnall, J.~Clarke, K.~Gudka, A.~Joannou, A.~T. Markettos, A.~Mazzinghi, R.~M. Norton, M.~Roe, P.~Sewell, S.~D. Son, T.~M. Jones, S.~W. Moore, P.~G. Neumann, and R.~N.~M. Watson, ``Cornucopia: Temporal safety for {CHERI} heaps,'' in \emph{2020 {IEEE} Symposium on Security and Privacy, {SP} 2020, San Francisco, CA, USA, May 18-21, 2020}.\hskip 1em plus 0.5em minus 0.4em\relax {IEEE}, 2020, pp. 608--625. [Online]. Available: \url{https://doi.org/10.1109/SP40000.2020.00098}
\BIBentrySTDinterwordspacing

\bibitem{cctag}
\BIBentryALTinterwordspacing
Z.~Liu, Y.~Rong, C.~Li, W.~Tan, Y.~Li, X.~Han, S.~Yang, and C.~Zhang, ``{CCTAG:} configurable and combinable tagged architecture,'' in \emph{32nd Annual Network and Distributed System Security Symposium, {NDSS} 2025, San Diego, California, USA, February 24-28, 2025}.\hskip 1em plus 0.5em minus 0.4em\relax The Internet Society, 2025. [Online]. Available: \url{https://www.ndss-symposium.org/ndss-paper/cctag-configurable-and-combinable-tagged-architecture/}
\BIBentrySTDinterwordspacing

\bibitem{devietti2008hardbound}
\BIBentryALTinterwordspacing
J.~Devietti, C.~Blundell, M.~M.~K. Martin, and S.~Zdancewic, ``Hardbound: architectural support for spatial safety of the {C} programming language,'' pp. 103--114, 2008. [Online]. Available: \url{https://doi.org/10.1145/1346281.1346295}
\BIBentrySTDinterwordspacing

\bibitem{DBLP:conf/osdi/ZeldovichKDK08}
\BIBentryALTinterwordspacing
N.~Zeldovich, H.~Kannan, M.~Dalton, and C.~Kozyrakis, ``Hardware enforcement of application security policies using tagged memory,'' in \emph{8th {USENIX} Symposium on Operating Systems Design and Implementation, {OSDI} 2008, December 8-10, 2008, San Diego, California, USA, Proceedings}, R.~Draves and R.~van Renesse, Eds.\hskip 1em plus 0.5em minus 0.4em\relax {USENIX} Association, 2008, pp. 225--240. [Online]. Available: \url{http://www.usenix.org/events/osdi08/tech/full\_papers/zeldovich/zeldovich.pdf}
\BIBentrySTDinterwordspacing

\bibitem{timberv}
\BIBentryALTinterwordspacing
S.~Weiser, M.~Werner, F.~Brasser, M.~Malenko, S.~Mangard, and A.~Sadeghi, ``{TIMBER-V:} tag-isolated memory bringing fine-grained enclaves to {RISC-V},'' in \emph{26th Annual Network and Distributed System Security Symposium, {NDSS} 2019, San Diego, California, USA, February 24-27, 2019}.\hskip 1em plus 0.5em minus 0.4em\relax The Internet Society, 2019. [Online]. Available: \url{https://www.ndss-symposium.org/ndss-paper/timber-v-tag-isolated-memory-bringing-fine-grained-enclaves-to-risc-v/}
\BIBentrySTDinterwordspacing

\bibitem{oracleadi}
``Application data integrity (adi),'' \url{https://docs.kernel.org/arch/sparc/adi.html}, [Online; accessed 1-April-2025].

\bibitem{califorms}
\BIBentryALTinterwordspacing
H.~Sasaki, M.~A. Arroyo, M.~T.~I. Ziad, K.~Bhat, K.~Sinha, and S.~Sethumadhavan, ``Practical byte-granular memory blacklisting using califorms,'' in \emph{Proceedings of the 52nd Annual {IEEE/ACM} International Symposium on Microarchitecture, {MICRO} 2019, Columbus, OH, USA, October 12-16, 2019}.\hskip 1em plus 0.5em minus 0.4em\relax {ACM}, 2019, pp. 558--571. [Online]. Available: \url{https://doi.org/10.1145/3352460.3358299}
\BIBentrySTDinterwordspacing

\bibitem{bogo}
\BIBentryALTinterwordspacing
T.~Zhang, D.~Lee, and C.~Jung, ``{BOGO:} buy spatial memory safety, get temporal memory safety (almost) free,'' in \emph{Proceedings of the Twenty-Fourth International Conference on Architectural Support for Programming Languages and Operating Systems, {ASPLOS} 2019, Providence, RI, USA, April 13-17, 2019}, I.~Bahar, M.~Herlihy, E.~Witchel, and A.~R. Lebeck, Eds.\hskip 1em plus 0.5em minus 0.4em\relax {ACM}, 2019, pp. 631--644. [Online]. Available: \url{https://doi.org/10.1145/3297858.3304017}
\BIBentrySTDinterwordspacing

\bibitem{intelmpx}
\BIBentryALTinterwordspacing
O.~Oleksenko, D.~Kuvaiskii, P.~Bhatotia, P.~Felber, and C.~Fetzer, ``Intel {MPX} explained: {A} cross-layer analysis of the intel {MPX} system stack,'' \emph{Proc. {ACM} Meas. Anal. Comput. Syst.}, vol.~2, no.~2, pp. 28:1--28:30, 2018. [Online]. Available: \url{https://doi.org/10.1145/3224423}
\BIBentrySTDinterwordspacing

\bibitem{safemem}
\BIBentryALTinterwordspacing
F.~Qin, S.~Lu, and Y.~Zhou, ``Safemem: Exploiting ecc-memory for detecting memory leaks and memory corruption during production runs,'' in \emph{11th International Conference on High-Performance Computer Architecture {(HPCA-11} 2005), 12-16 February 2005, San Francisco, CA, {USA}}.\hskip 1em plus 0.5em minus 0.4em\relax {IEEE} Computer Society, 2005, pp. 291--302. [Online]. Available: \url{https://doi.org/10.1109/HPCA.2005.29}
\BIBentrySTDinterwordspacing

\bibitem{rest}
\BIBentryALTinterwordspacing
K.~Sinha and S.~Sethumadhavan, ``Practical memory safety with {REST},'' in \emph{45th {ACM/IEEE} Annual International Symposium on Computer Architecture, {ISCA} 2018, Los Angeles, CA, USA, June 1-6, 2018}, M.~Annavaram, T.~M. Pinkston, and B.~Falsafi, Eds.\hskip 1em plus 0.5em minus 0.4em\relax {IEEE} Computer Society, 2018, pp. 600--611. [Online]. Available: \url{https://doi.org/10.1109/ISCA.2018.00056}
\BIBentrySTDinterwordspacing

\bibitem{libmpk}
\BIBentryALTinterwordspacing
S.~Park, S.~Lee, W.~Xu, H.~Moon, and T.~Kim, ``libmpk: Software abstraction for intel memory protection keys (intel {MPK}),'' in \emph{2019 USENIX Annual Technical Conference (USENIX ATC 19)}.\hskip 1em plus 0.5em minus 0.4em\relax Renton, WA: USENIX Association, Jul. 2019, pp. 241--254. [Online]. Available: \url{https://www.usenix.org/conference/atc19/presentation/park-soyeon}
\BIBentrySTDinterwordspacing

\bibitem{song2024parallel}
\BIBentryALTinterwordspacing
W.~Song, D.~Xie, Z.~Xue, and P.~Liu, ``A parallel tag cache for hardware managed tagged memory in multicore processors,'' \emph{IEEE Transactions on Computers}, vol.~73, no.~11, pp. 2488--2503, 2024. [Online]. Available: \url{https://ieeexplore.ieee.org/document/10633901}
\BIBentrySTDinterwordspacing

\end{thebibliography}

\newpage %

\appendices %

\revision{
\section{Meta-Review}

The following meta-review was prepared by the program committee for the 2026
IEEE Symposium on Security and Privacy (S\&P) as part of the review process as
detailed in the call for papers.

\subsection{Summary}

The paper presents \sys, a system that improves the bug-detection capability of ARM’s Memory Tagging Extension (MTE) by enabling byte-granular detection of heap memory safety bugs in unmodified binaries. The authors first conduct an empirical study on real MTE hardware (Google Pixel 8) and show that MTE’s fixed 16-byte tag granularity causes it to miss a substantial fraction of heap-based buffer overflows, particularly intra-granule overflows. They demonstrate that such cases are common in real programs and can lead to silent failures even for real-world vulnerabilities. To address this limitation, \sys combines MTE’s low-overhead hardware checks with selective software-based checking using a sampling-based “tripwire” mechanism for short (partially used) tag granules.

\subsection{Scientific Contributions}
\begin{itemize}
\item Creates a New Tool to Enable Future Science.
\item Provides a Valuable Step Forward in an Established Field.
\end{itemize}

\subsection{Reasons for Acceptance}
\begin{enumerate}
\item This paper provides a valuable step forward in an established field. Overall, the paper provides a practical engineering improvement that increases detection granularity while maintaining relatively low overhead, advancing the feasibility of fine-grained memory error detection on real hardware and motivating further research on hybrid hardware/software memory safety techniques. The paper shows that ARM MTE's coarse granularity can cause a meaningful portion of heap overflows to go unnoticed, making a good case for exploring byte-level detection. The proposed approach can increase the detection granularity while maintaining manageable overheads.
\item The paper creates a new tool to enable future science. A real prototype of the proposed idea has been implemented and evaluated.
\end{enumerate}

\subsection{Noteworthy Concerns} %
\begin{enumerate} %
\item The paper does not have a threat model or a discussion of the security implications of sampling. This makes it unclear if NanoTag would be applicable to production systems. For example attackers in an adversarial setting could wait for the end of the slow start to attack.
\item The paper does not discuss the trade-offs of the probabilistic approach in a fuzzing environment, where a probabilistic technique will make it hard to reproduce bugs.
\end{enumerate}

}

\end{document}